\documentstyle[11pt,aaspp4]{article}
\begin{document}
\newcommand{\lya}{Lyman~$\alpha$}
\newcommand{\lyb}{Lyman~$\beta$}
\newcommand{\degpoint}{\mbox{$^\circ\mskip-7.0mu.\,$}}
\newcommand{\minpoint}{\mbox{$'\mskip-4.7mu.\mskip0.8mu$}}
\newcommand{\secpoint}{\mbox{$''\mskip-7.6mu.\,$}}
\newcommand{\sqdeg}{\mbox{${\rm deg}^2$}}
\newcommand{\squig}{\sim\!\!}
\newcommand{\subsun}{\mbox{$_{\twelvesy\odot}$}}
\newcommand{\et}{{\it et al.}~}
\newcommand{\Rs}{{\cal R}}
\newcommand{\iras}{{\it IRAS\/}}

\def\ltsima{$\; \buildrel < \over \sim \;$}
\def\simlt{\lower.5ex\hbox{\ltsima}}
\def\gtsima{$\; \buildrel > \over \sim \;$}
\def\simgt{\lower.5ex\hbox{\gtsima}}
\def\propsima{$\; \buildrel \propto \over \sim \;$}
\def\simprop{\lower.5ex\hbox{\propsima}}
\def\arcs{$''~$}
\def\arcm{$'~$}
\title{MULTI-WAVELENGTH OBSERVATIONS OF DUSTY STAR FORMATION
AT LOW AND HIGH REDSHIFT}

\author{\sc Kurt L. Adelberger and Charles C. Steidel\altaffilmark{1,2}}
\affil{Palomar Observatory, Caltech 105--24, Pasadena, CA 91125}

\altaffiltext{1}{NSF Young Investigator}
\altaffiltext{2}{Packard Fellow}

\begin{abstract}
If high-redshift galaxies resemble rapidly star-forming galaxies
in the local universe, most of the luminosity produced
by their massive stars will have been absorbed by dust and
re-radiated as far-infrared photons that cannot
be detected with existing facilities.
This paper examines what can be
learned about high-redshift star formation from the
small fraction of high-redshift galaxies' luminosities
that is emitted at accessible wavelengths.  We first consider
the most basic ingredient in the analysis
of high-redshift surveys: the estimation of star-formation
rates for detected galaxies.  
Standard techniques require an
estimate of the bolometric luminosity produced
by their massive stars. 
We review and quantify empirical correlations between
bolometric luminosities produced by star formation and the UV, mid-IR, sub-mm, and radio
luminosities of galaxies in the local universe.
These correlations suggest that observations of high-redshift galaxies
at any of these wavelengths should
constrain their star-formation rates to within $\sim$ 0.2--0.3 dex.
We assemble the limited evidence that high-redshift
galaxies obey these locally calibrated correlations.
The second part of the paper assesses whether existing surveys
have found the galaxies that host the majority of star formation at high redshift
even though they directly detect only a small fraction of the luminosities
of individual galaxies.  We describe the characteristic
luminosities and dust obscurations of galaxies at
$z\sim 0$, $z\sim 1$, and $z\sim 3$.  After discussing the relationship
between the high-redshift populations selected in surveys
at different wavelengths, we calculate the contribution
to the $850\mu$m background from each and argue that
these known galaxy populations
can together have produced the entire
observed background.  
The available data show that a correlation between
star-formation rate and dust obscuration $L_{\rm bol,dust}/L_{\rm UV}$
exists at low and high redshift alike.  The existence of
this correlation plays a central role in the
major conclusion of this paper:  most star formation
at high redshift occurred in galaxies
with moderate dust obscurations $1\simlt L_{\rm bol,dust}/L_{\rm UV}\simlt 100$
similar to those that host the majority of star formation in
the local universe and to those that are detected in UV-selected
surveys. 

\end{abstract}
\newpage

\section{INTRODUCTION}

The universe of galaxies beyond $z\sim 1$ has finally become
relatively easy to observe.  While only a handful of galaxies
at $z\simgt 1$ were known five years ago, close to
2000 have spectroscopic redshifts today.  Planned redshift
surveys should increase the number by at least an
order of magnitude in the next several years.
Analyzing the large and rapidly growing samples
of high-redshift galaxies will be a major challenge
in the coming decade.

Much of this analysis will rely on estimates of
star-formation rates for the detected galaxies.  
These estimates are required to answer the most
basic questions that high-redshift surveys can address---when did the stars in
the universe form?  how were the sites of
star formation related to the evolving
perturbations in the underlying distribution
of matter?---but unfortunately no standard and widely
accepted techniques exist for estimating star-formation
rates from the data usually available
in high-redshift samples.  Researchers analyzing data taken
at a variety of wavelengths with a variety of
techniques have reached wildly discrepant conclusions about
the star-formation history of the universe
(e.g. Madau \et 1996; Smail \et 1997; Barger \et 1998;
Blain \et 1999a; Blain \et 1999c; Eales \et 1999; Pettini \et 1999;
Steidel \et 1999; Tan, Silk, \& Balland 1999; Trentham, Blain, \& Goldader 1999;
Barger, Cowie, \& Richards 2000, hereafter BCR).

The central problem is that the energy produced by star formation
in young galaxies is emitted across more than 6 decades
of frequency, from the far-UV to the radio,
while most high-redshift surveys are based on observations
at only a narrow range of frequencies.  Standard methods
for deriving a galaxy's star-formation rate require an estimate
of the total energy emitted by its massive stars, but
in existing surveys this is never directly observed.
Radio surveys detect only the tiny percentage of the total
energy that goes into accelerating electrons in supernova remnants,
for example, while optical surveys detect the portion of
massive stars' emission that is not absorbed by dust
and sub-mm surveys detect a fraction of the portion that is.
Large and uncertain corrections are needed to estimate the total
emitted energy from the energy detected at any of these
wavelengths.  Differences in the adopted corrections are
largely responsible for current discrepancies in estimates
of the star-formation history of the universe.

In principle one could obtain a robust estimate of the energy
emitted by massive stars in a high redshift galaxy
by observing it over a sufficiently wide range of wavelengths,
but with current technology this is rarely possible.  Star-forming
galaxies at high redshift almost certainly emit the bulk
of their luminosities at far-infrared wavelengths that
are extremely difficult to detect.  Although massive stars
themselves radiate mainly in the far-UV, rapid star formation
in the local universe occurs in dusty environments where
most UV photons are quickly absorbed, and as a result
most of the luminosity produced by massive stars 
tends to emerge in the far-IR where
dust radiates.  This is known to be true for a broad range of
galaxies in the local universe, from 
the famous class of ``Ultra Luminous Infrared
Galaxies'' (ULIRGs, galaxies with $L_{\rm FIR}\ga 10^{12} L_\odot$)
to the much fainter UV-selected starbursts
contained in the IUE Atlas (e.g. Meurer, Heckman, \& Calzetti 1999,
hereafter MHC).
The recent detection of a large extragalactic far-IR
background (e.g. Fixsen et al. 1998) suggests that it
is likely to have been true at high redshifts as well.

Because most of the energy produced by massive stars is likely
radiated by dust in the far-IR, estimates of the bolometric
dust luminosities of high redshift galaxies are required
in order to estimate their star-formation rates.  The bolometric
dust luminosities usually cannot be directly observed---the
earth's atmosphere is opaque in the far-IR and most high-redshift 
galaxies are too faint to have been detected
by any far-IR satellite that has flown---but they can be inferred
from known correlations between local galaxies' bolometric
dust luminosities and their luminosities in the optical, mid-IR,
sub-mm, and radio atmospheric windows.

The first part of this paper aims to present a unified framework for estimating
the bolometric dust luminosities of high-redshift galaxies
selected at different wavelengths.  In \S 2 we summarize
and re-examine locally established correlations between
bolometric dust luminosities and luminosities through
atmospheric windows, quantifying
their scatter in cases where it has not been done explicitly
by previous authors.  In \S 3 we present the limited available
evidence that high-redshift galaxies obey these correlations.
Published estimates of the cosmic star-formation history
derived from observations through one atmospheric
window have often been accompanied by claims
that most of the star formation could not have
been detected with observations through other atmospheric windows.
In the second part of this paper, \S 4, we use the
results of \S\S 2 and 3 to evaluate whether this is true.
After reviewing the properties of star-forming galaxies
in the local universe, we discuss the properties of
high-redshift galaxies selected in surveys
at different wavelengths and estimate the contribution
to the far-IR background from galaxies similar
to those detected in optical, mid-IR, and sub-mm surveys.
Our conclusions are presented in \S 5.

Throughout this paper we shall denote luminosities
by $L$ when they have units of energy time$^{-1}$, by $l_\nu$
when they have units of energy time$^{-1}$ frequency$^{-1}$, 
and by $l_\lambda$ when they have units of energy time$^{-1}$ wavelength$^{-1}$;
and observed fluxes by $f_\nu$ (units of energy time$^{-1}$ frequency${-1}$ area$^{-1}$)
or $f_\lambda$ (units of energy time$^{-1}$ wavelength$^{-1}$ area$^{-1}$).

\section{DUST-OBSCURED STAR FORMATION IN THE LOCAL UNIVERSE}
The energy produced by massive stars emerges from galaxies 
over a wide range of wavelengths.  Figure 1 shows
typical spectral shapes of this emission for rapidly star-forming
galaxies in the local universe.
Massive stars radiate predominantly in
the far-UV, but as galaxies become
dustier an ever larger fraction of their luminosity
is absorbed by dust and re-radiated in the far-IR.
In typical cases most of the energy released by massive
stars emerges in the far-IR.  This section describes
how the dust's bolometric luminosity can be estimated
from observations through the atmospheric
windows indicated on the $x$-axis of Figure 1.

\subsection{Sub-mm constraints on bolometric dust luminosity}
Although a galaxy's total dust luminosity at 
$10\mu{\rm m}\simlt\lambda\simlt 1000\mu{\rm m}$ cannot be
directly measured from the ground, narrow atmospheric
windows in the sub-millimeter (Figure 1) allow
a small fraction of it to reach the earth's
surface.  If dust in all galaxies radiated with the
same known spectral energy distribution (SED)---say a 50K blackbody---then
a galaxy's bolometric dust luminosity could be reliably estimated
by simply scaling the flux measured through a sub-mm atmospheric 
window.\footnote{provided, of course, that the galaxy's
redshift and luminosity distance are known.}
Unfortunately dust SEDs vary considerably
from one galaxy to the next and are more complex
than simple blackbodies.
Obtaining a reasonable fit to a galaxy's measured dust SED
usually requires at least two ``modified'' blackbodies of the form 
$$f_\nu \propto {\nu^{3+\epsilon}\over \exp({h\nu / kT})-1},\eqno(1)$$
where $\epsilon\sim 1.5$, the emissivity index, is sometimes treated
as a free parameter in addition to the temperatures and 
amplitudes of the modified
blackbodies (e.g. Klaas \et 1997).
The required parameters in these fits vary among galaxies
in a way that is not quantitatively understood.

Because of the significant variation in the shape of
galaxies' dust SEDs, measuring a galaxy's flux in
one of the narrow sub-mm atmospheric windows cannot
precisely determine its bolometric dust luminosity.  Nevertheless, sub-mm fluxes
provide some useful information; galaxies
with large fluxes through the sub-mm atmospheric windows
tend also to have large bolometric dust luminosities.  
To quantify how well sub-mm flux measurements can
constrain the bolometric dust luminosities of actively
star-forming galaxies in the local universe, we have
assembled from the literature a sample of these galaxies
with published \iras\ fluxes at $25\mu$m, $60\mu$m, and $100\mu$m,
as well as at least one flux measurement at a wavelength 
$200\mu{\rm m}<\lambda<1000\mu{\rm m}$ to constrain
the temperature of the cool dust component and the slope of
its (modified) Rayleigh-Jeans tail.  This sample
consists of 27 galaxies:
the ULIRGs IRAS05189-2524, IRAS08572+3915, UGC5101,
IRAS12112+0305, Mrk231, Mrk273, IRAS14348-1447,
and IRAS15250+3609 detected at $800\mu$m (and sometimes $450\mu$m) by
Rigopoulou \et (1996); the LIRGs NGC1614, NGC3110,
NGC4194, NGC4418, NGC5135, NGC5256, NGC5653, NGC5936, Arp193,
and Zw049 detected at $850\mu$m
by Lisenfeld \et 1999; the interacting galaxies NGC6240, Arp220,
and Arp244 observed at $25\mu{\rm m} < \lambda < 200\mu{\rm m}$
by Klaas \et (1997); the UV-selected starbursts
NGC6090, NGC7673, NGC5860, IC1586, and Tol1924-416
detected at $150\mu$m and $205\mu$m by Calzetti \et (1999); 
and the nearby starburst
M82 (Hughes, Robson, \& Gear 1990 and references therein).

The dust SEDs of these galaxies are shown in Figure 2.
For clarity we have connected measurements of individual galaxies
at different wavelengths with spline fits (continuous lines).  These
fits were calculated in $\log\lambda - \log f_\lambda$ space
and should roughly capture the expected power-law shape
of dust SEDs in their (modified) Rayleigh-Jeans tails.  Measurements
from all galaxies were normalized to have the same
total luminosity under the fits.  

This plot
can be used to estimate a rapidly star-forming galaxy's bolometric
dust luminosity on the basis of its luminosity at a single
observed wavelength:
$$L_{\rm bol,dust} \sim \biggl\langle{L_{\rm bol,dust}\over\lambda l_\lambda}\biggr\rangle \lambda l_\lambda \equiv {\cal K}(\lambda) \lambda l_\lambda \eqno(2)$$
where the expectation value is taken over all dust SEDs in the
figure.  Table 1 lists the values ${\cal K}$ at various
wavelengths.  Ideally one would calculate ${\cal K}$
by fitting two modified blackbodies to each dust SED, as discussed above,
but most of the galaxies in our sample have measured fluxes
at only 4 relevant wavelengths ($25\mu$m, $60\mu$m, $100\mu$m, and $850\mu$m)
and this fit has more than 4 free parameters.
Instead, we estimated ${\cal K}$ from the log-log spline interpolations
shown in Figure 2.  This spline fitting does not
take proper account of uncertainties in the measured fluxes, but
the uncertainties are almost always far smaller than the
intrinsic differences in dust SEDs among the galaxies in our
sample.  The observed spread among these galaxies
in $L_{\rm bol,dust} / \lambda l_\lambda$ at a given wavelength
provides a measure of how reliably a flux measurement at that
wavelength can constrain a galaxy's bolometric dust luminosity.
The RMS spread in $\log_{10}{\cal K}$, also listed in Table 1,
suggests sub-mm estimates of $L_{\rm bol,dust}$ will be
accurate to about a factor of 2.

According to Sanders \& Mirabel (1996) and Dunne \et (2000), 
the shapes of galaxies'
dust SEDs are correlated with their bolometric luminosities,
and so one might hope to estimate $L_{\rm bol,dust}$ more
accurately by using different values of ${\cal K}$ for
galaxies with different absolute sub-mm luminosities.
Unfortunately this trend does not
appear to be strong enough among rapidly star-forming
galaxies to substantially reduce the scatter
in ${\cal K}$ (i.e. to substantially improve estimates of
$L_{\rm bol,dust}$ derived from photometry in a single sub-mm window).
Illustrating this point, Figure 3 shows
${\cal K}(200\mu{\rm m})$ ($\equiv L_{\rm bol,dust}/\lambda l_\lambda$ at
$\lambda=200\mu$m) versus $L_{\rm bol,dust}$ for each galaxy in our
sample.  Although LIRGs clearly have systematically different
${\cal K}(200\mu{\rm m})$ values than ULIRGs, the trend disappears at
lower luminosities.  Including an absolute luminosity
dependence in ${\cal K}(\lambda)$ seems unlikely
to significantly improve sub-mm constraints on $L_{\rm bol, dust}$.

Measuring the luminosity of a galaxy in two sub-mm bands
instead of only one provides additional
information about the shape of the dust SED that could
in principle be used to estimate $L_{\rm bol,dust}$ more accurately.
Unfortunately the major sub-mm atmospheric windows
at $450\mu$m and $850\mu$m both lie in the Rayleigh-Jeans
tail of typical dust SEDs at $z\simlt 3$.
Measuring both $450\mu$m and $850\mu$m fluxes of a galaxy
at these redshifts therefore
reveals little about the temperature of the cold dust component
and almost nothing at all about the possible presence of other
dust components at higher temperatures.
At larger redshifts measuring both fluxes provides better
constraints on the shape of the dust SED, but even then
two flux measurements cannot uniquely determine the
shape of dust SEDs that typically require more than four
free parameters for a reasonable fit.
In any case accurate fluxes in more than one sub-mm band are
seldom available for the faint high-redshift galaxies that are our
primary concern.  Obtaining these additional sub-mm fluxes
may ultimately prove useful for more accurate estimation
of galaxies' bolometric dust luminosities, but we will not consider
this possibility further.

\subsection{Mid-Infrared constraints on bolometric dust luminosity}
At wavelengths $\lambda\simlt 15\mu$m the dust emission
from galaxies tends to rise significantly above the
modified blackbody fits discussed in \S 2.1.  
This excess emission is thought to originate from very small grains transiently
heated by single UV-photons to high non-equilibrium temperatures;
much of it emerges in a few discrete polycyclic aromatic
hydrocarbon (PAH) lines at
3.3, 6.2, 7.7, 8.6, and $11.3\mu$m (see, e.g., Xu \et 1998
and references therein).  $7.7\mu$m is often the
strongest of these lines and the easiest to 
detect at high redshift.  In this subsection
we will attempt to quantify the constraints that
observations through a 6--9$\mu$m filter can
place on a galaxy's bolometric dust luminosity.
We will need this information in \S 3.3 below.

The sample we will use consists of 11 starbursts
and 53 star-formation--dominated ULIRGs
observed in the mid-IR by Rigopoulou \et (2000)
and the starburst NGC6090 observed in the mid-IR by Acosta-Pulido \et (1996).

In analogy to \S 2.1 we will define a constant of proportionality
${\cal K}_{\rm MIR}$ between bolometric dust luminosity
and luminosity at $6\mu{\rm m}<\lambda<9\mu{\rm m}$:
$$L_{\rm bol, dust} = {\cal K}_{\rm MIR} 
	\int_{6\mu{\rm m}}^{9\mu{\rm m}}d\lambda\, l_\lambda.\eqno(3)$$
After excluding two outliers with anomalously
low $\sim 8\mu$m luminosities, IRAS00199-7426 and IRAS20100-4156,
the mean value of ${\cal K}_{\rm MIR}$ among the galaxies
in this sample is 52, the mean value of $1/{\cal K}_{\rm MIR}$
is 0.025, and the rms dispersion in 
$\log_{10}{\cal K}_{\rm MIR}$ is 0.22 dex.
Each galaxy's bolometric dust luminosity was estimated from its
measured IRAS $60\mu$m and $100\mu$m luminosities
with the relationship $L_{\rm bol,dust}\simeq 1.47L_{FIR}$, where
$L_{FIR}\equiv 0.65 L_{60} + 0.42 L_{100}$,
$L_{60}\equiv\nu l_\nu$ at $\nu=c/60\mu$m, and
$L_{100}\equiv\nu l_\nu$ at $\nu=c/100\mu$m
(e.g., Helou \et 1988).  The adopted relationship
between $L_{\rm bol,dust}$ and $L_{FIR}$
is satisfied by the galaxies in the sample of \S 2.1
with an rms scatter of 7\%.  
The estimated mean and rms of ${\cal K}_{\rm MIR}$
are not significantly altered by restricting
our analysis either to galaxies with flux measurements
in the sub-mm as well as at 60 and $100\mu$m
or to the galaxies with the least noisy mid-IR spectra.
Similar values of ${\cal K}_{\rm MIR}$ are observed
in galaxies with a wide range of bolometric luminosities,
as shown in Figure 4.

\subsection{UV constraints on bolometric dust luminosity}
The methods of estimating $L_{\rm bol,dust}$ discussed so far
rely on fitting an average dust SED through the observed
dust emission at a single wavelength.  $L_{\rm bol,dust}$ can also
be estimated without observing any dust emission at all.
Many standard methods of estimating the dust content of
galaxies rely on the absorption signature of dust in
the UV/optical rather than its emission in the far-IR.
We will focus our attention on one of these methods, which
has been shown (see, e.g., MHC)
to predict reasonably well the observed dust emission
of local starburst galaxies.

The idea behind this technique is simple.  Because dust
absorbs shorter wavelengths more strongly than longer wavelengths,
the UV continua of starbursts should become redder on average as the galaxies
become dustier and the ratio of far-IR to far-UV luminosity
increases.  According to MHC, there is surprisingly little
scatter in this trend among starburst galaxies in
the local universe; 
the bolometric dust luminosities of the 57 starburst galaxies
in their sample obey
$$L_{\rm bol,dust} = 1.66 (10^{0.4(4.43+1.99\beta )}-1) L_{1600},\eqno(4)$$
where $L_{1600} \equiv \nu l_\nu$ evaluated at $\nu=c/1600$\AA\
and $\beta$ is the observed spectral slope ($l_\lambda\propto\lambda^\beta$)
at $1200 \simlt \lambda \simlt 2000$\AA, 
with a scatter of 0.3 dex.  This correlation is illustrated
in Figure 1.  

\subsection{Radio constraints on bolometric dust luminosity}
Observations through the radio atmospheric window can also
constrain $L_{\rm bol,dust}$.  Increases in far-IR luminosity
are accompanied, in local galaxies, by increases in radio emission
with a typical spectral slope at $\nu\sim c/20{\rm cm}$
of $f_\nu\propto\nu^{-0.8}$ (Condon 1992).
The radio emission is thought to be synchrotron radiation
from electrons in supernova remnants; it is proportional
to $L_{\rm bol,dust}$ presumably because supernovae occur
in the same population of massive stars responsible for heating the dust.
According to Condon (1992), galaxies without active nuclei
in the local universe satisfy
$$L_{\rm bol,dust} \simeq 7.9\times 10^5 L_{20}, \eqno(5)$$
where $L_{20}\equiv \nu l_\nu$ at $\nu=c/20{\rm cm}$, with a scatter
of 0.2 dex.  This relationship is usually expressed in terms
of the quantity $L_{FIR}$;
the constant of proportionality
in eq. 5 assumes $L_{\rm bol,dust}\simeq 1.47L_{FIR}$,
as discussed in \S 2.2. 

The local radio/far-IR relation (eq. 5) is unlikely to hold
to arbitrarily high redshifts.  At some redshift
the energy density of the cosmic microwave background,
increasing as $(1+z)^4$, will become comparable
to the magnetic energy density in star-forming galaxies, and
relativistic electrons in supernova
remnants will begin to cool through inverse Compton scattering
rather than synchrotron emission.
For magnetic energy densities comparable to those
observed among rapidly star-forming galaxies in
the local universe, this will occur
at $z\simgt$ 3--6 (e.g. Carilli \& Yun 1999).

We conclude this section with a brief digression.
Carilli \& Yun (1999) have recently proposed
the ratio $\eta\equiv f_\nu(850\mu{\rm m}) / f_\nu(20{\rm cm})$
as a photometric redshift estimator.
Results in this section allow us to
quantify the estimator's likely accuracy.
Equations 2 and 5 imply
$$\eta\simeq {3400 (1+z)^{-1-\alpha} \over {\cal K}(850\mu{\rm m}/1+z)},\eqno(6)$$
where $\alpha\simeq -0.8$ is the spectral index of the radio emission
and ${\cal K}(\lambda)$ can be interpolated
from table 1. 
Calculating the mean expected $\eta$ and its uncertainty
as a function of redshift
is complicated by the logarithmic uncertainties and by
the presence of ${\cal K}$ in the denominator of eq. 6.
To calculate the mean expected value of $\eta$ we used
$\bar\eta = 3400 (1+z)^{-\alpha-1} \langle 1/{\cal K}(850\mu{\rm m}/1+z)\rangle,$
where $\langle 1/{\cal K}\rangle$ is the mean value of
$1/{\cal K}$ among the galaxies in \S 2.1; to calculate
the (asymmetrical) $1\sigma$ uncertainties in $\eta$ we assumed that
$\log_{10}(\eta)$ was symmetrically distributed around
$\mu \equiv \log_{10}(\bar\eta) - \sigma_{\rm dex}^2 \log_e{10} / 2$,
where $\sigma_{\rm dex}$,
the logarithmic spread in $\eta$ for galaxies at a given redshift,
is the quadrature sum of the logarithmic
uncertainty in ${\cal K}(\lambda)$ (from Table 1)
and the 0.2 dex scatter in the radio/far-IR correlation.  If the uncertainty in
$\log_{10}(\eta)$ has a Gaussian distribution---as is roughly
the case---this choice of $\mu$ correctly reproduces the
desired mean $\bar\eta$.
We will handle denominator terms and logarithmic uncertainties
similarly throughout this paper unless stated otherwise.

Figure 5 shows the expected range of $\eta$ for star-forming
galaxies as a function of redshift; redshift has been placed on
the ordinate to allow redshift constraints to be easily read
off this figure given a measured value of $\eta$.
This figure can be used to estimate redshifts of a star-forming
galaxies on the basis of their $850\mu$m and $20$cm fluxes,
but the estimates will be accurate
only if these galaxies obey the local radio to far-IR correlation (eq. 5)
and have similar dust SEDs as the local sample of \S 2.1.  Our next
section is concerned, in part, with whether this is true (see also
Carilli \& Yun 2000).

\section{OBSERVATIONS AT HIGH REDSHIFT}

With very few exceptions, galaxies at high redshift
are too faint to have been detected over the
wavelength range $20\mu{\rm m}\simlt\lambda\simlt 200\mu{\rm m}$
by any satellite that has flown.  Their bolometric
dust luminosities therefore cannot be directly measured;
they can only be indirectly estimated with the correlations we have just
described between $L_{\rm bol,dust}$ and luminosities 
at accessible wavelengths.  Unfortunately the
underlying physics responsible for these
empirical correlations is poorly understood,
and so there is no particular theoretical
reason to expect them to exist in identical form at high-redshift.
Nor can their existence be directly established empirically,
for without measuring high-redshift galaxies' fluxes
at several wavelengths $20\mu{\rm m}\simlt\lambda\simlt 200\mu{\rm m}$
it is impossible to show directly
that their bolometric dust luminosities are correlated in the expected way
with their fluxes through various atmospheric windows.  It might therefore
appear that the assumptions required to estimate high-redshift galaxies'
bolometric dust luminosities (and therefore star-formation rates)
are not only questionable but also untestable; but this is not
entirely true.  If a galaxy's fluxes in the rest-UV, mid-IR, sub-mm,
and radio are each correlated with its bolometric dust
luminosity, then they should also be correlated
with each other in a way that is straightforward
to calculate.  Observing high-redshift galaxies at more
than one of these wavelengths therefore provides
an indirect test of standard methods for estimating
$L_{\rm bol,dust}$.  These indirect tests are the
subject of this section.

Because our ultimate goal is to estimate the $850\mu$m fluxes
of known UV-selected high-redshift populations from
their UV colors, our emphasis in this section will
be on testing whether high-redshift galaxies have
the mid-IR, sub-mm, and radio fluxes that their UV-colors
and the correlations in \S 2 would predict.  This
should not give the false impression that we are
testing solely whether MHC's $\beta$/far-IR relation
is satisfied by high-redshift galaxies.
Without directly measured bolometric dust luminosities,
it is impossible to test for the existence of any one
of the correlations of \S 2 independently of the others; they can only
be tested in pairs.  If $z\sim 3$ galaxies
are found to have radio fluxes at odds with
our predictions from their far-UV properties, for example,
it might mean that they do not obey the local $\beta$/far-IR
relation, but it could equally well mean that they
do not obey the local radio/far-IR relation.
What we are really testing in this section is whether
high-redshift star-forming galaxies are sufficiently similar to
their low redshift counterparts for their
bolometric dust luminosities and star-formation rates
to be reliably estimated in the absence of complete
photometry at rest wavelengths $20\mu{\rm m}\simlt\lambda\simlt 200\mu{\rm m}$.

The observations required for these tests are extremely difficult with current
technology, and even though the data analyzed in this
section are among the best available, much of the evidence
presented is rather marginal.  We have included
even inconclusive evidence below when it
illustrates what can (and cannot) be learned
from currently feasible observations.

\subsection{SMMJ14011}
The lensed galaxy SMMJ14011+0252 at 
$\alpha({\rm J2000})=14^{\rm h}01^{\rm m}04^{\rm s}\mskip-5.6mu .97$, 
$\delta({\rm J2000})=+02^o52'24''\mskip-7.6mu .6$, 
$z = 2.565$ (Smail \et 1998, Frayer \et 1999) is the only sub-mm--selected object
robustly confirmed to be a high redshift galaxy 
without an obviously active nucleus (e.g. Ivison \et 2000).
$450\mu$m, $850\mu$m, and 20cm fluxes
for this object were obtained by Ivison \et (2000).
We obtained rest-UV photometry through the $G$ and ${\cal R}$ filters
(Steidel \& Hamilton 1993)
with COSMIC (Kells \et 1998) on the Palomar 200 inch Hale telescope
in March 1999.  Our optical photometry of SMMJ14011 is presented
in Table 2.

Predicting SMMJ14011's radio and sub-mm fluxes
from its UV photometry requires an estimate of the slope
of its UV continuum, $\beta$ (in
$f_\lambda\propto\lambda^\beta$; see \S 2.3).
Estimating $\beta$ from UV photometry is not
entirely trivial for two reasons.  First,
the observed UV continua of high-redshift galaxies
have been reddened by the intergalactic absorption of the Lyman-$\alpha$
forest, and this must be corrected before we can
estimate their intrinsic continuum slopes $\beta$.
Second, the UV continua of starburst galaxies are not exactly
power laws, and so fitting their continua over different
wavelength ranges can produce different estimates
of the best fit slope $\beta$.
MHC calculated $\beta$ for galaxies in their local sample
by fitting these galaxies' fluxes
through ten ``windows'' spanning the wavelength
range $1260{\rm\AA} < \lambda < 2600{\rm\AA}$; nine of these
windows were bluer than 1950\AA.  
The available photometry of high-redshift galaxies
rarely covers an identical wavelength range.
At SMMJ14011's redshift of $z=2.565$, for example,
the $G$ and ${\cal R}$ filters sample $\sim 1350$\AA\ 
and $\sim 1950$\AA\ rest.
To estimate $\beta$ for SMMJ14011 in a way that
took some account of these complications,
we first subjected
a GISSEL96 model of an actively star-forming galaxy (Bruzual \& Charlot 1996)
to the appropriate intergalactic absorption (Madau 1995), and then
reddened the resulting spectrum with increasing dust extinction following
a Calzetti (1997) law until this synthetic spectrum had the same
$G$ and ${\cal R}$ colors as SMMJ14011.  Finally we fit this synthetic
spectrum through the 10 windows of MHC with a power-law
of the form $l_\lambda\propto\lambda^\beta$.  The best fit $\beta$
we find---which is rather insensitive to the details of this procedure---is
$\beta=-0.74$, with a $1\sigma$ uncertainty due to photometric
errors of $\pm 0.25$.  

Equations 2--4 can now be used to predict SMMJ14011's
radio and sub-mm fluxes from its UV photometry.
Defining the constants
$${\cal K}_{850}(z) \equiv L_{\rm bol,dust} / \nu l_\nu 
\quad{\rm at}\quad \nu = c(1+z)/850\mu{\rm m},\eqno(7a)$$
$${\cal K}_{450}(z) \equiv L_{\rm bol,dust} / \nu l_\nu
\quad{\rm at}\quad \nu = c(1+z)/450\mu{\rm m},\eqno(7b)$$
$${\cal K}_L(z,\alpha) \equiv L_{\rm bol,dust} / \nu l_\nu\simeq 7.9\times 10^5 (1+z)^{-(1+\alpha)}
\quad{\rm at}\quad \nu = c(1+z)/20{\rm cm},\eqno(7c)$$
and
$${\cal K}_{\cal R}(z,\beta) \equiv L_{1600} / \nu l_\nu\simeq (4.3/1+z)^{\beta+1}
\quad{\rm at}\quad \nu = c(1+z)/6900{\rm\AA},\eqno(7d)$$
where $\alpha\simeq -0.85\pm 0.16$ (Richards 2000) is
the slope of the synchrotron emission (in $f_\nu\propto\nu^\alpha$),
$\beta$ is the slope of the UV-continuum, as defined above,
and 6900\AA\ is approximately the
central wavelength of the ${\cal R}$ filter, and using
$f_\nu = 10^{-0.4(m_{\rm AB}+48.60)}$ ergs/s/cm$^2$/Hz, 
we find, after substituting into equations 2, 4, and 5, and
changing from $l_\nu$ to $f_\nu$ with the
equation $l_{\nu_1}/l_{\nu_2} = f_{\nu_1}/f_{\nu_2}$, the following
relationships between the observed-frame optical, sub-mm, and radio
fluxes of star-forming galaxies at $z\sim 3$:
$$f_{\nu, {\rm obs}}(850\mu{\rm m}) \sim 10^{-0.4({\cal R}-22.17)}
\biggl({{\cal K}_{\cal R}(z,\beta) \over 1.}\biggr)
\biggl({{\cal K}_{850}(z) \over 10.}\biggr)^{-1}
(10^{0.4(4.43+1.99\beta )}-1)\quad {\rm mJy} \eqno(8)$$
$$f_{\nu, {\rm obs}}(450\mu{\rm m}) \sim 10^{-0.4({\cal R}-22.98)}
\biggl({{\cal K}_{\cal R}(z,\beta) \over 1.}\biggr)
\biggl({{\cal K}_{450}(z) \over 2.5}\biggr)^{-1}
(10^{0.4(4.43+1.99\beta )}-1)\quad {\rm mJy} \eqno(9)$$
$$f_{\nu, {\rm obs}}(1.4{\rm GHz}) \sim 10^{-0.4({\cal R}-23.86)}
\biggl({{\cal K}_{\cal R}(z,\beta) \over 1.}\biggr)
\biggl({{\cal K}_L(z,\alpha) \over 6.0\times 10^5}\biggr)^{-1}
(10^{0.4(4.43+1.99\beta )}-1)\quad \mu{\rm Jy} \eqno(10)$$
with no dependence on cosmology.
The redshift dependence is contained in the $K$-corrections
${\cal K}_{450}$, ${\cal K}_{850}$, ${\cal K}_{\cal R}$, and ${\cal K}_L$;
the default values of ${\cal K}_{850}=10$,
${\cal K}_{450}=2.5$, ${\cal K}_{\cal R}=1$, ${\cal K}_L=6.0\times 10^5$
in this equation, derived from equation 7 and Table 1, are 
roughly appropriate for $z\simeq 3.0$. 
The uncertainty in these equations is large.  
Eq. 4 is able to predict $L_{\rm bol,dust}$ from starbursts'
UV luminosities and spectral slopes with an uncertainty of
about 0.3 dex.  At fixed $L_{\rm bol,dust}$ there is (as argued
in \S 2.1) an additional uncertainty of 0.1 dex in $l_\nu$ 
at $100\mu{\rm m}\simeq 450\mu{\rm m}/1+z$, 0.3 dex
at $200\mu{\rm m}\simeq 850\mu{\rm m}/1+z$, and
0.2 dex at $6{\rm cm}\simeq 20{\rm cm}/1+z$.
If these logarithmic uncertainties add in quadrature, we would expect
equations 8, 9, and 10 to be able to predict sub-mm and radio
fluxes from UV fluxes to within 0.3--0.4 dex,
if high-redshift galaxies resembled local starbursts exactly.  

For a galaxy at $z=2.565$ with $\beta=-0.74$, the appropriate
$K$-correction constants for equations 8--10 are
${\cal K}_{\cal R}=1.05$,
${\cal K}_L=6.13\times 10^5$,
and, interpolating between the values listed in Table 1,
${\cal K}_{850}=15.3$ and
${\cal K}_{450}=2.6$.
These equations then predict $850\mu{\rm m}$, $450\mu{\rm m}$, 20cm fluxes for SMMJ14011
of $23^{+14}_{-9}$ mJy, $70^{+43}_{-28}$ mJy, $150^{+95}_{-67} \mu$Jy.
The quoted uncertainties reflect only
the uncertainty in $\beta$ from photometric errors.  An additional
uncertainty of $\sim 0.4$ dex applies to each predicted
flux, and so these predicted fluxes
agree well with the measured fluxes (Ivison \et 2000)
$f_\nu(850\mu{\rm m}) = 15\pm 2$mJy,
$f_\nu(450\mu{\rm m}) = 42\pm 7$mJy,
$f_\nu(20{\rm cm}) = 115\pm 30\mu$Jy (see Figure 6).
SMMJ14011 evidently exhibits the same relationship between its
UV, far-IR, and radio properties as rapidly star-forming
galaxies in the local universe.

\subsection{Lyman-break Galaxies}
\subsubsection{Sub-mm}
Although it is encouraging that one galaxy at $z\sim 3$, SMMJ14011+0252,
appears similar to local galaxies in the correlations of
its rest-frame UV, sub-mm, and radio fluxes, ideally we would
like to establish that this holds true for entire
populations of galaxies at $z\sim 3$.  Unfortunately this is exceedingly
difficult with current technology.  None of the $\sim 800$
spectroscopically confirmed $z\sim 3$ galaxies in the
UV-selected sample of Steidel \et (2000), for example,
has an ${\cal R}$ magnitude
as bright as SMMJ14011's ${\cal R}=21.25$, and $\sim$ 95\% of
the galaxies in that sample are more than ten times fainter.  
Even if these galaxies were as heavily obscured as SMMJ14011,
we would expect their $850\mu$m fluxes to be at least ten
times lower than SMMJ14011's, or $\simlt 1$mJy, well below
SCUBA's detection limit.  (SCUBA, on the James Clerk Maxwell
Telescope, is the current state-of-the-art sub-mm bolometer
array; see Holland \et 1999).  This suggests that (typical) individual
Lyman-break galaxies are unlikely to be detected
with SCUBA, and tests of whether these galaxies have the
bolometric dust luminosities predicted from their UV spectral slopes
will be at best statistical.

The faintness of typical Lyman-break galaxies makes their
sub-mm and radio fluxes difficult to predict as well.
Predicted bolometric dust luminosities depend 
sensitively on the UV spectral slope $\beta$, and
this slope is difficult to estimate accurately for
faint objects with relatively large photometric errors.  For example,
if a galaxy at $z=2.6$ had $G-{\cal R}=0.75$, we would
estimate (using the method of \S 3.1) $\beta=-0.60$ and
$L_{\rm bol,dust} \simeq 31\times L_{1600}$, while
if it had $G-{\cal R}=0.95$ we would estimate $\beta=-0.07$
and $L_{\rm bol,dust}= 85\times L_{1600}$.
A photometric error of 0.2 in $G-{\cal R}$ is not unusual
for Lyman-break galaxies, especially at fainter magnitudes 
(e.g. Adelberger \et 2000), and so photometric errors
can easily affect our predicted sub-mm and radio fluxes by
a factor of $\sim 3$.  

The situation is worsened by the fact that
the only Lyman-break galaxies likely to be
detectable in the sub-mm or radio are those with
reddest UV slopes and highest dust obscurations.  
Sub-samples of Lyman-break galaxies selected
for follow-up observations at these wavelengths
will have to be chosen from among the reddest
members of the population.  Because highly reddened
Lyman-break galaxies are much rarer than their
less reddened counterparts (cf. Figure 12, discussed below), and
because the typical photometric errors on Lyman-break
galaxies are large, there is a reasonable
chance that a Lyman-break galaxy which appears
to be heavily reddened is actually a less reddened galaxy
with photometric errors.  These sub-samples
will therefore suffer from significant Malmquist bias.

For these reasons, observing Lyman-break galaxies in the sub-mm 
does not appear to be an especially promising way
to test whether these galaxies obey the $\beta$/far-IR correlation.
Nevertheless Chapman \et (2000) have attempted
the observations.  Their results are shown in Figure 7.

In order to calculate predicted sub-mm fluxes for these
galaxies in a way that accounted
for the significant photometric uncertainty and Malmquist bias
discussed above, and for the expected intrinsic scatter in
the UV/$850\mu$m relationship (eq. 8), we proceeded as
follows.  We first generated a large ensemble of simulated Lyman-break galaxies
with ${\cal R}$ magnitudes drawn at random from the $z\sim 3$
LBG luminosity function shown in Figure 12, UV slopes $\beta$
drawn at random from the distribution in Figure 12 (cf. Steidel \et (1999),
Adelberger \et (2000)), and redshift $z$ drawn at random from
the interval $2.5<z<3.5$.  Each of these galaxies was assigned
a $G$ magnitude by inverting the procedure of \S 3.1, and
an $850\mu$m flux using eq. 8.  The $850\mu$m fluxes were
then changed randomly to reflect the scatter
in the $\beta$/far-IR relation and in ${\cal K}_{850}(z)$.
Finally, we placed each simulated galaxy
at a random location in our $G$ and ${\cal R}$ images and attempted
to detect it and measure its photometry with the same
software used to find real Lyman-break galaxies in these images.
This produced a long list of the ${\cal R}$ magnitudes
and $G-{\cal R}$ colors we would expect to measure for Lyman-break galaxies
with known redshifts and $850\mu$m fluxes if the
galaxies obeyed the local $\beta$/far-IR relation and
had dust SEDs similar their local analogs.  In order
to predict the $850\mu$m fluxes shown in Figure 7,
we binned together the $850\mu$m fluxes of simulated galaxies with magnitudes,
colors, and redshifts similar to those of each galaxy in the sample
of Chapman \et (2000).  As our best guess of the predicted
$850\mu$m flux for each galaxy in the Chapman \et sample,
we took the sample mean of the $850\mu$m
flux among simulated galaxies in its bin; for the uncertainty
in the predicted flux we took the standard deviation
of the $850\mu$m fluxes of these same simulated galaxies.

For simplicity we have indicated in Figure 7 the uncertainties
in the predicted fluxes with the convention
mean $\pm$ rms, but in fact these uncertainties
are very skewed.  The Monte Carlo simulations just described
suggest that most Lyman-break galaxies should have $850\mu$m
fluxes lower than the mean value and that some will have
$850\mu$m fluxes significantly larger (how else
could we have mean $\sim$ rms with the predicted
fluxes bounded below by zero?), and so the observations
are more consistent with the predictions than Figure 7
might suggest at first glance.

One way to quantify the agreement of the observations with
our predictions is to calculate $\chi^2$.
Denoting the predicted fluxes by $x$ and the observed by $y$,
the relevant $\chi^2$ figure of merit,
$$\chi^2\equiv\sum_{i=1}^N {(x_i-y_i)^2 \over \sigma_{x,i}^2 + \sigma_{y,i}^2}\eqno(11)$$
(e.g. Press \et (1992, \S 15.3)) is equal to 14.5.
There are 11 data points and no free parameters in this fit,
and so the data are at least not grossly inconsistent with
our expectations.  It is difficult to make a more quantitative
statement with this approach, however, because the
error bars on the predicted flux are very skewed,
and $\chi^2$ is therefore not drawn from the
usual distribution derived for Gaussian uncertainties.

An alternate approach is to ask what these data say about
slope $b$ of the line $f_{\rm observed} = b f_{\rm predicted}$.
If Lyman-break galaxies obeyed the $\beta$/far-IR correlation
and had dust SEDs similar to those of local galaxies, we
would expect the data to be consistent with $b=1$.
A Bayesian approach can quantify the constraints
on $b$ in a way that takes proper account of our skewed
uncertainties.  This approach requires us to calculate
the likelihood of the data as a function of $b$.  In the
brief derivation below,
$f_m$ and $\sigma$ denote measured $850\mu$m fluxes and uncertainties
for the galaxies in Chapman \et (2000),
$f_t$ denotes these galaxies' (unknown) true fluxes that would
be measured if we had perfect data,
$f_t'\equiv f_t/b$ denotes the true $850\mu$m fluxes
of the {\it simulated} Lyman-break galaxies, calculated above 
under the assumption $b=1$,
and~$I$ is shorthand for the background knowledge
(such as the luminosity and $\beta$ distributions of Lyman-break
galaxies, and the scatter in the $\beta$/far-IR relation, and so on)
that went into our Monte Carlo simulations.
Our data set consists of the redshifts $\{z\}$, apparent magnitudes
$\{\Rs\}$, colors $\{G-\Rs\}$, and $850\mu$m fluxes $\{f_m\}$
of each galaxy the sample of Chapman \et (2000).
The likelihood of observing these data given $b$ is equal to
the product of the likelihoods of observing each galaxy
individually:
$$P\Bigr(\{G-\Rs\}\{\Rs\}\{z\}\{f_m\}\,|\,\{\sigma\}bI\Bigl) =
\prod_i^N P\Bigr((G-\Rs)_i\Rs_iz_if_{m,i}\,|\,\sigma_i bI\Bigl).\eqno(12)$$
These individual likelihoods can be evaluated by integrating
over the unknown true $850\mu$m flux of each galaxy:
$$P\Bigl((G-\Rs)\Rs z f_m\,|\,\sigma bI\Bigr) =
\int_0^{\infty}df_t\ P\Bigl(f_t\,|\,bI\Bigr)
P\Bigl((G-\Rs)\Rs z\,|\,f_t I\Bigr)
P\Bigl(f_m\,|\,f_t\sigma\Bigr).\eqno(13)$$
We have simplified this equation by omitting irrelevant
variables from the conditional probabilities and
by writing $P(AB)$ as $P(A)P(B)$ when $A$ and $B$ are independent.
$P\Bigl(f_m\,|\,f_t\sigma\Bigr)$ can be calculated by assuming
that the uncertainties in the measured fluxes are
Gaussian, which is roughly true.
The remaining probabilities in the integral can be estimated
from the Monte Carlo simulations.  Writing the total number of
galaxies in the simulations as $N_{\rm tot}$, and the
number of galaxies in the simulations with both properties
$A$ and $B$ as $n(A,B)$, we have
$P\Bigl(f_t\,|\, bI\Bigr)\simeq n(f_t')/N_{\rm tot}$ and
$P\Bigl((G-\Rs)\Rs z\,|\,f_t I\Bigr)\simeq n(G-\Rs,\Rs,z,f_t')/n(f_t')$.
Substituting into eq. 12 yields the likelihood, given $b$, of observing
a single galaxy at redshift $z$ with UV photometry $\Rs$ and $G-\Rs$
and $850\mu$m flux $f_m$:
$$P\Bigl((G-\Rs)\Rs z f_m\,|\,\sigma bI\Bigr) \propto
\int_0^{\infty}df_t'\ n(G-\Rs,\Rs,z,f_t')\,
\exp\biggl[-{1\over 2}\biggl({bf_t'-f_m\over\sigma}\biggr)^2\biggr].\eqno(14)$$

The maximum likelihood value of $b$, calculated numerically
using equations 12 and 14, is $b=0.45$.
For a uniform prior, the 68\% and 90\% highest posterior density
regions are $0.22\leq b\leq 0.87$ and $0.09\leq b\leq 1.2$.
The sub-mm fluxes of Lyman-break galaxies
therefore appear to be somewhat lower than we would
have expected.  This could indicate that
Lyman-break galaxies do not quite satisfy
MHC's $\beta$/far-IR correlation, or that their dust SEDs are 
different from those of the local galaxies described
in \S 2.1; but the inconsistency with $b=1$ is only marginally
significant and we will not speculate further about
its possible cause.   All we can say with much
confidence is that the
sub-mm fluxes of Lyman-break galaxies
are unlikely to be either much higher or more than an order of
magnitude lower than our predictions.  

Our interpretation of these data differs quantitatively from that
of Chapman \et (2000).  This is because those authors
ignored the uncertainties in the predicted fluxes in most
of their analysis, and assumed that the dust SEDs of 
all Lyman-break galaxies would be well approximated by a single
modified black-body (eq. 1) with $T=50$K and $\epsilon=1.5$, rather
than by the range of empirical SEDs estimated in \S 2.1.

\subsubsection{Radio}

Radio observations of Lyman-break galaxies can in principle provide
another indirect test of standard methods for estimating
these galaxies' bolometric dust luminosities.  
Extremely deep radio images are required, however,
if we hope to put interesting limits on the radio
fluxes of typical $z\sim 3$ galaxies:
equation 10 implies that the vast majority of Lyman-break
galaxies in the sample of Steidel \et (2000)
will have 20cm fluxes of only a few $\mu$Jy
or less.  One of the deepest existing radio images is Richards' (2000)
20cm VLA map of the Hubble Deep and Flanking Fields, which
reaches a $1\sigma$ noise limit of $8\mu$Jy/beam.  
Unfortunately even this image does not appear sufficient
for a meaningful test.  Of the 46 Lyman-break
galaxies in our spectroscopic sample that lie within this
20cm image, only 2 are predicted (using the Monte-Carlo
method of \S 3.2.1) to have marginally
detectable fluxes $f_\nu>10\mu$Jy.  These two, with predicted
fluxes of $12\pm 15\mu$Jy and $19\pm 17\mu$Jy, have
observed fluxes of $4\pm 12\mu$Jy and $-6\pm 12\mu$Jy---not
exactly in line with our predictions, but not damning either.
(Our method of estimating the observed fluxes is described in
\S 3.3.2 below.)  Nine galaxies have
marginally significant observed fluxes of $f_\nu>10\mu$Jy,
and none of these nine were predicted to be especially
bright; but six galaxies have measured fluxes $f_\nu<-10\mu$Jy,
and so even the marginally significant detections
may simply reflect the noise characteristics of Richards' (2000)
image.  Statistical tests based on the sample as a whole
are as inconclusive as the object-by-object tests we have
just described:  the total observed and predicted 20cm fluxes from
these 46 galaxies are $105\pm 81\mu$Jy and $114\pm 36\mu$Jy, respectively,
and the formal 90\% credible interval on 
$b\equiv f_{\rm observed}/f_{\rm predicted}$, estimated
using the Monte Carlo simulations and Bayesian approach
of \S 3.2.1, is $0<b<2.1$.

\subsection{Balmer-break galaxies}
Galaxies at redshifts as high as $z\sim 3$ are hardly ideal for testing
whether the UV-properties of high-redshift galaxies can
be used to predict their bolometric dust luminosities,
because large luminosity distances and unfavorable
$K$-corrections make them extremely faint in the
optical, mid-IR, and radio (cf. Figure 1).
A better redshift is $z\sim 1$.  At this redshift
galaxies are much brighter in these atmospheric windows,
and their far-UV spectral slopes $\beta$ can still
be measured with ground-based photometry.
In this section we will use galaxies drawn from the
$z\sim 1$ ``Balmer-break'' sample of Adelberger \et (2000)
to test whether actively star-forming galaxies at $z\sim 1$
follow the correlations between UV, mid-IR, and radio
properties observed among galaxies in the local universe.
This color-selected sample consists of $\sim 700$ star-forming
galaxies with spectroscopic redshifts $0.9\simlt z\simlt 1.1$
and apparent magnitudes ${\cal R}\simlt 25.5$.

One warning is due before we begin.  The $\beta$/far-IR correlation
of MHC (eq. 4) has been shown to hold only for starburst
galaxies in the local universe.  It is not known whether
less rapidly star-forming galaxies such as the Milky Way
satisfy this correlation, and post-starburst
galaxies, whose red UV slopes reflect age rather
than dust content, almost certainly do not.
At $z\sim 3$ galaxies are
bright enough to satisfy the Lyman-break criteria of
Steidel \et (1999) only if they are forming stars rapidly
($\simgt 30 M_\odot$/yr with typical dust obscurations,
for $h=0.7$, $\Omega_M=0.3$, $\Omega_\Lambda=0.7$),
and so these galaxies might reasonably be expected
to obey the $\beta$/far-IR correlation (see MHC for
further discussion of the Lyman-break galaxy/starburst
connection).  But at $z\sim 1$ post-starburst galaxies
and slowly star-forming galaxies like the Milky Way
can both satisfy our Balmer-break selection criteria,
in addition to the starburst galaxies that are expected
to obey the $\beta$/far-IR correlation.  There is therefore
good reason to suspect {\it a priori} that some galaxies
in our Balmer-break sample will not obey this correlation.

In principle one might be able to select the starburst analogs
from among a Balmer-break sample on the basis of their
photometric or spectroscopic signatures, but a proper
treatment is beyond the scope of this paper.
Fortunately slowly star-forming galaxies will be far
too faint in the radio and mid-IR to be detected
in current data, and so their presence in our Balmer-break
sample will not significantly affect the conclusions
of this subsection.
A crude way of partially
excluding post-starbursts, which we shall adopt below, is to impose
an additional selection criterion $U_n-G<0.5$ on the
Balmer-break sample.  This limits the Balmer-break
sample to the same range of $\beta$ observed among
starburst galaxies in the local universe and among
Lyman-break galaxies at $z\sim 3$ 
by excluding the galaxies that
are reddest in the far-UV, whose slopes might
reflect age rather than their dust content. 
We will discuss this further elsewhere (Adelberger \et 2000).
 
\subsubsection{Mid-Infrared}
Because of the strength of the $7.7\mu$m PAH feature,
$15\mu$m observations provide a particularly promising
way of observing dust emission from galaxies at $z\sim 1$.
Seventy-one Balmer-break galaxies from our spectroscopic sample
lie within the $15\mu$m (LW3) ISO image
of the CFRS 1415+52 field obtained by Flores \et (1999);
six of them are listed as optical counterparts
to $15\mu$m sources in these authors' $3\sigma$ catalog.

Equations 3 and 4 imply the following relationship
between the UV and mid-IR photometry of galaxies at $z\sim 1$:
$$f_{\nu,{\rm obs}}(15\mu{\rm m}) \sim 10^{-0.4(U_n-25.22)}
    \biggl({{\cal K}_U(z,\beta) \over 1}\biggr)
    \biggl({{\cal K}_{15} \over 20}\biggr)^{-1}
    (10^{0.4(4.43+1.99\beta)}-1)\quad \mu{\rm Jy},\eqno(15)$$
where 
$${\cal K}_{15} \equiv L_{\rm bol, dust} / \nu \bar l_\nu 
	\quad{\rm at}\quad \nu = c(1+z)/15\mu{\rm m}, \eqno(16)$$
$${\cal K}_U(z,\beta) \equiv L_{1600} / \nu l_\nu 
	\simeq \biggl({1+z\over 2.3}\biggr)^{\beta+1}
	\quad{\rm at}\quad \nu = c(1+z)/3700{\rm \AA}, \eqno(17)$$
$\bar l_\nu$ is the mean luminosity density at
rest frequencies $c(1+z)/18\mu{\rm m} < \nu < c(1+z)/12\mu{\rm m}$,
and 3700\AA\ is roughly the central wavelength of the
$U_n$ filter after accounting for atmospheric absorption and
instrumental throughput.
In most cases the observed $15\mu$m fluxes
of $z\sim 1$ galaxies are dominated by emission
from the 7.7 and $8.6\mu$m PAH lines, and ${\cal K}_{15}$
depends very weakly on redshift near $z\sim 1$.
It is well approximated by ${\cal K}_{\rm MIR}/2.4\simeq 22$
for all the redshifts we will consider here (${\cal K}_{\rm MIR}$
is defined in \S 2.2). 
 
To estimate $\beta$ for these galaxies we used their
$U_n-G$ colors, which sample rest wavelengths $\sim 1850$\AA\
and $\sim 2350$\AA, and a procedure analogous to the one
described in \S 3.1.
Photometric errors are less of a concern for these
optically bright galaxies than for the faint
$z\sim 3$ galaxies of \S 3.2.  Instead of taking the elaborate
Monte-Carlo approach of \S 3.2, we estimated
the predicted $15\mu$m fluxes for these galaxies 
by assuming that their true $U_n$ magnitudes
and $U_n-G$ colors were equal on average to the measured
values; for the uncertainty in the predicted fluxes
we simply propagated through the uncertainties in the
$U_n-G$ and $U_n$ fluxes determined from the Monte-Carlo
simulations.  The total uncertainty in the predicted
fluxes, 0.5 dex, arises from the scatter
in the $\beta$/far-IR correlation (0.3 dex), from the
expected scatter in ${\cal K}_{15}$ (0.2 dex),
from photometric uncertainties in $U_n$ magnitudes (0.1 dex),
and from the uncertainty in $\beta$ due to photometric
errors in $U_n-G$ (0.3 dex).

Figure 8 shows the predicted and observed $15\mu$m fluxes of these 71 
Balmer-break galaxies.  The majority of these galaxies
do not appear in the $3\sigma$ catalog of Flores \et (2000),
and so we know only that their fluxes were less than $\sim 150\mu$Jy;
these sources are indicated with downward pointing arrows.
The detections are indicated by points with error bars.

These data appear encouragingly consistent with our UV-based
predictions:  the large number of galaxies with predicted
flux significantly lower than the detection limit are not 
detected, while the majority of those with predicted
fluxes close to (or above) the detection limit are.
The only apparently serious discrepancy between predicted and
observed fluxes (ISO138 with $f_{\rm predicted}\sim 4$,
$f_{\rm observed}\sim 250$), may result from a misidentification
of the optical galaxy responsible for the $15\mu$m emission;
according to Flores \et (2000)
there is a 27\% chance of misidentification for this
object, compared to a median 3\% chance for the
other five detections.

Readers who do not find Figure 8 persuasive may wish to consider
the following.  The seven galaxies that we predicted to be brightest
have the following ranks in observed flux: 5, 6, $>$6, 3, $>$6, 1, 4,
where undetected objects are assigned the rank ``$>$6'' because
we know only that they are fainter than the six detected objects.
Over 90\% of the Balmer-break galaxies in the $15\mu$m image
have the observed rank ``$>$6''; 
if our UV-predicted $15\mu$m fluxes were unrelated
to the observed fluxes, we would have expected $\sim 90$\% of the
seven objects predicted to be brightest to have the rank ``$>$6'', but
that is not the case.
This type of reasoning can be quantified
with a version of Kendall's $\tau$ generalized for
censored data sets by Oakes (1982):  of 500 simulated data sets
we have made by scrambling the observed rank assigned
to each predicted rank, only 2 had
a generalized $\tau$ as large or larger than the real data set.
The predicted and observed fluxes of these galaxies
evidently display a significant positive correlation;
and moreover the fact that the handful of detections
have observed fluxes broadly in line with their
predicted fluxes suggests that this 
correlation is consistent with the expected linear correlation 
$\langle f_{\rm predicted}\rangle = \langle f_{\rm observed}\rangle$.

\subsubsection{Radio}
Eighty-nine of the Balmer-break galaxies from the spectroscopic 
sample of Adelberger \et (2000) lie within the deep 1.4GHz
image of the Hubble Deep Field recently published
by Richards (2000).  
We estimated rough 1.4GHz fluxes
for this sample by summing the flux within a 2\secpoint0
radius aperture surrounding the optical centroid of each object.
The measured fluxes are shown
in Figure 9.  For the $1\sigma$ uncertainty we took
$\sim 12\mu$Jy, the observed rms when we
estimated fluxes in a similar manner for random locations
in Richards' (2000) image.  The predicted fluxes
were estimated using
$$f_{\nu,{\rm obs}}(20{\rm cm}) \sim 10^{-0.4(U_n-24.19)}
    \biggl({{\cal K}_U(z,\beta) \over 1}\biggr)
    \biggl({{\cal K}_L \over 6.9\times 10^5}\biggr)^{-1}
    (10^{0.4(4.43+1.99\beta)}-1)\quad \mu{\rm Jy},\eqno(18)$$
which follows from equations 4, 5, 7c, and 17. 
Otherwise the predicted radio fluxes were derived
in an identical manner to the predicted mid-IR fluxes
in \S 3.3.1.
The uncertainty in the predicted fluxes arises from the scatter
in the $\beta$/far-IR correlation (0.3 dex), from the
scatter in the far-IR/radio correlation (0.2 dex),
from photometric uncertainties in $U_n$ magnitudes (0.1 dex),
and from the uncertainty in $\beta$ due to photometric
errors in $U_n-G$ (0.4 dex).  Added in quadrature these
amount to a $\sim 0.5$ dex uncertainty in the predicted
20cm flux for each object.  

Because of the large uncertainties in the predicted and
observed fluxes, these data can hardly provide a rigorous
test of whether $z\sim 1$ galaxies obey the correlations
between far-UV, far-IR, and radio properties observed
in rapidly star-forming galaxies in the local universe.
However, despite the unimpressive appearance of Figure 9,
the data do strongly suggest that the predicted
and observed fluxes of Balmer-break galaxies are at least
positively correlated:  fewer than 1\% of simulated data sets
generated by randomly shuffling the observed
and predicted fluxes had a Kendall's $\tau$ as large
as the observed data set.  The trend of higher
measured fluxes for objects with higher predicted fluxes
is illustrated by the dotted crosses in Figure 9, which
show the mean measured flux ($\pm$ the standard deviation
of the mean) for objects in different bins of predicted
flux.  Although objects with higher predicted
fluxes clearly tend to have higher observed fluxes,
it appears that the measured 20cm fluxes may 
be somewhat lower than our predictions, especially
for the brightest objects.  The significance of this
result is difficult to assess without running
Monte-Carlo simulations similar to those
discussed in \S 3.2.  These simulations are not yet available.

\subsection{HR10}
The $z=1.44$ galaxy HR10 initially attracted the attention of 
Hu \& Ridgway (1994)
because of its unusually red optical-to-infrared color.
Those authors suggested that HR10 was a high-redshift
elliptical galaxy, but subsequent observations
provided hints that HR10 was actively forming stars
and suggested that its red colors might result
from extreme dust reddening rather than from
an aged stellar population (Graham \& Dey 1996)
This interpretation has received strong support from
the recent detection of HR10 at $450\mu$m and $850\mu$m.
HR10 is now considered the prototypical example of an
extremely dusty star-forming galaxy at high redshift.
Its photometry is presented in Figure 10 (Dey \et 1999,
Haynes \et 2000). 

Because of the large uncertainties in its rest-frame
far-UV colors, we cannot reliably measure $\beta$ for
HR10, but we can use its observed ratio
of far-UV to far-IR luminosity to estimate the value of
$\beta$ required for this object to satisfy MHC's $\beta$/far-IR
relation.  This slope, shown in Figure 10, is not obviously
inconsistent with the data.

Although it is unclear if HR10 satisfies the $\beta$/far-IR relation,
we can say with confidence that its ratio of far-IR to UV luminosity
far exceeds that of any other galaxy discussed
in this section.  Despite the large star-formation rate implied
by its $850\mu$m flux, HR10 would not be included in most
optically selected high-redshift surveys.  Placed at $z=3.0$, for example,
HR10 would be too faint and too red to satisfy commonly used  ``Lyman-break''
photometric selection criteria (e.g. Steidel, Pettini, \& Hamilton 1995).
If most of the star formation in the universe occurred in objects
like HR10, rather than in the relatively UV-bright
Balmer-break and Lyman-break galaxies previously discussed,
then UV-selected high-redshift surveys will
provide a seriously incomplete view of star formation
in the universe.  But this seems an unlikely possibility
to us, as we will now explain.

\section{UV-SELECTED POPULATIONS, SUB-MM SOURCES, AND THE FAR-IR BACKGROUND}

In \S 3 we tried to assess whether the star-formation rates of known
high-redshift galaxies can be estimated from their
fluxes through various atmospheric windows.  Our conclusions
were uncertain but hopeful:  the meager available evidence
is at least consistent with the idea that ground-based
observations can be used together with locally calibrated
correlations to provide rough constraints on 
high-redshift galaxies' star-formation rates.
But estimating moderately accurate star-formation rates
is only one step towards the ultimate goal of high-redshift
surveys, producing a reasonably complete census
of star formation at high redshift that can be used
to constrain our empirical and theoretical understanding
of the early stages of galaxy formation.  
In this section we will set aside the object-by-object
comparisons of star-formation rates estimated from
photometry at different wavelengths, and turn our
attention instead to the larger question of
whether existing surveys are approaching the
reasonably complete view of high-redshift
star formation that we desire.

There is good reason to be concerned that the view of high-redshift star formation
provided by these surveys is far from complete.  Few
star-forming galaxies in the local universe would be included in
existing surveys if they were placed at high redshift.
This is illustrated by Figure 11a, which
shows the ratio of dust to ultraviolet luminosity for local galaxies
with various star-formation rates.  For this plot we have adopted
$L_{1600}+L_{\rm bol,dust}$ as a crude measure of star-formation
rate, though in some cases 1600\AA\ luminosities and dust luminosities
can be powered by processes other than star formation.  Because of
the different constants of proportionality in the
conversions of $L_{1600}$ and $L_{\rm bol,dust}$ to
star-formation rates, the quantity $1.66 L_{1600} + L_{\rm bol,dust}$
(cf. eq. 4; see also MHC) would likely provide a better measure
of star-formation rate, but in almost all cases
$L_{1600}$ is small compared to $L_{\rm bol,dust}$
and our adoption of the simpler $L_{1600}+L_{\rm bol,dust}$
will not appreciably affect the subsequent discussion.

Solid circles on Figure 11a denote UV-selected starbursts from the
sample of MHC, discussed earlier in the text; \S\ symbols denote
nearby spirals\footnote{NGC247, NGC1232, NGC1313, NGC1398, NGC2403, NGC2683, NGC2976, NGC3556, 
NGC3623, NGC3726, NGC3877, NGC4178, NGC4216, NGC4254, NGC4303, NGC4307, NGC4321, 
NGC4522, NGC4559, NGC4565, NGC4592, NGC4595, NGC4653, NGC5055, NGC5248, NGC5457, 
NGC5806, NGC5879, NGC5907, NGC6946}, selected
randomly from among the objects present in both the Carnegie Atlas of Sandage \& Bedke (1994)
and the far-UV catalog of Rifatto, Lango, \& Capaccioli (1995);
open squares denote three LIRGs (NGC4793, NGC5256, and NGC6090)
from the catalog of Sanders, Scoville, \& Soifer (1991) which
had UV fluxes estimated by Kinney \et (1993) or Rifatto \et (1995);
and open stars denote the ULIRGs VII Zw 031, IRAS F12112+0305, and IRAS F22691-1808
observed in the far-UV by Trentham, Kormendy, \& Sanders (1999).
Bolometric dust luminosities were estimated from these galaxies'
$60\mu$m and $100\mu$m IRAS fluxes as discussed in \S 2.4.  The adopted
conversion from IRAS fluxes to bolometric dust luminosities is
more appropriate for the starbursts, LIRGs, and ULIRGs in Figure 11a
than for the spirals, but the resulting errors in the spirals' estimated
dust luminosities will not affect our conclusions significantly.

Taken together the galaxies in this plot are representative of
those that host the majority of star formation in the local universe:
according to Heckman (1998), starbursts and spirals account for 
roughly equal shares of the bulk of star formation in the local universe,
while LIRGs and ULIRGs (galaxies with $L_{FIR}\simgt 10^{11}L_\odot$)
account for perhaps 6\% (Sanders \& Mirabel 1996).

The curved lines on this plot show the $z=3.0$ completeness
limits ($\Omega_M=0.3$, $\Omega_\Lambda=0.7$) for existing surveys at optical (solid line),
sub-mm (dashed line), and radio (dotted line) wavelengths.
These lines assume an optical flux limit of 
$f_\nu(1600{\rm\AA}\times (1+z))\simgt 0.2\mu$Jy (${\cal R}_{\rm AB}\simlt 25.5$),
roughly appropriate for the Lyman-break survey
of Steidel \et (1999), a sub-mm flux limit of $f_\nu(850\mu{\rm m})\simgt$ 2mJy,
the confusion limit of SCUBA on the JCMT, and a radio
flux limit of $f_\nu(20{\rm cm})\simgt 50\mu$Jy, the completeness
limit of Richards' (2000) catalog of HDF sources.
Observations with lower flux limits have been obtained in each
of these bands, especially the optical---the flux limit in the Hubble
Deep Field is nearly an order of magnitude fainter than
the $0.2\mu$Jy optical limit we have adopted---but the
flux limits above are roughly appropriate for the
bulk of existing ``deep'' observations at each wavelength.
Sub-mm and radio flux limits were converted into dust luminosity limits
under the assumption that high-redshift galaxies
follow the empirical correlations described in \S 2.
The minimum luminosity for sub-mm detection is roughly independent of redshift,
because $K$-corrections largely cancel out changes in luminosity distance,
but the minimum luminosity for optical and radio detection would
be (respectively) 7.4 and 13 times fainter at $z\sim 1$ (cf. Figures 11b and 11c),
and 2.2 and 3.1 times brighter at $z\sim 5$.
Evidently only a small fraction of star-forming galaxies
in the local universe would be detected in existing surveys
if they were placed at $1\simlt z\simlt 5$.  

Nevertheless large numbers of high-redshift galaxies have been found.
Figure 11b shows the location on the star-formation rate versus
dust obscuration plot of the majority of known galaxies at $z\sim 1$.
Solid circles represent optically selected Balmer-break galaxies at $0.8<z<1.2$ from
the sample of Adelberger \et (2000), open stars represent $15\mu{\rm m}$
ISO sources at $0.8<z<1.2$ from the HDF sample discussed in \S 4.2 below, open
squares (triangles) represent radio and sub-mm sources with estimated
redshifts $z<2$ from the optically detected (undetected) sample of BCR,
and the open circle represents HR10.  Each object's dust luminosity was
estimated from its available ground-based photometry through
the empirical correlations described in \S 2.  In cases where
an object's dust luminosity could have been estimated with
more than one of the empirical correlations, we took an
estimate from the object's flux in only one wavelength range,
with the following order of preference: $850\mu{\rm m}$,
$15\mu{\rm m}$, $20{\rm cm}$, UV.
The dust luminosities of ISO sources in the HDF, for example, were
estimated from their $15\mu{\rm m}$ fluxes (\S 2.2) rather than from
their UV spectral slopes (\S 2.3) or radio fluxes (\S 2.4).
The curved lines in Figure 11b show typical $z=1.0$ detection thresholds for
observations at different wavelengths.  The optical, sub-mm, and radio
thresholds assume the flux limits discussed above; the mid-IR limit
assumes a $15\mu{\rm m}$ flux limit of $75\mu$Jy, which is
roughly appropriate for the HDF catalog of Aussel \et (1999).  
Because the objects in this plot do not lie at precisely $z=1.0$,
and because small amounts of data exist with
flux limits deeper than the typical values adopted here, 
some detected objects have star-formation luminosities $L_{1800}+L_{\rm bol,dust}$
lower than the indicated thresholds.

Figure 11c is a similar plot for galaxies at $z\sim 3$.  The solid
circles represent optically selected Lyman-break galaxies from the sample of 
Steidel \et (2000), open circles represent the sub-mm sources
SMMJ14011+0252 and west MMD11, the open squares (triangles) represent
radio/sub-mm sources with estimated redshifts $z>2$ from the optically
detected (undetected) sample of BCR, and the solid triangle
is the optically undetected sub-mm source SMMJ00266+1708
with an estimated redshift $z\sim 3.5$ (Frayer \et 2000).
In this plot (and in Figure 17 below) the gravitational
lensing magnification of SMMJ14011+0252 and SMMJ00266+1708
was roughly corrected by dividing their observed
luminosities by 2.75 (Frayer \et 1999) and 2.4 (Frayer \et 2000) respectively.

At each of these redshifts, $z\sim 0$, $z\sim 1$, and $z\sim 3$, star-forming
galaxies appear to have a similar range of dust obscuration
$0.1\simlt L_{\rm bol,dust}/L_{\rm UV} \simlt 1000$, and
to follow a similar correlation of dust obscuration and
star-formation rate,
but the bolometric star-formation luminosities of detected
galaxies increase steadily from $z\sim 0$ to $z\sim 1$ to $z\sim 3$.
This is partly a selection effect, but it also reflects
a genuine increase in the number density of rapidly
star-forming galaxies at higher redshifts (e.g. Lilly \et 1996, Steidel \et 1999,
BCR).  Although existing high-redshift surveys
are not deep enough (for the most part) to detect star-forming
galaxies similar to those in the local universe, the observed increase
in bolometric star-formation luminosity with redshift
suggests that these surveys may nevertheless have
detected a substantial fraction of high-redshift star formation.

The goal of this section is to estimate how large a fraction
of high-redshift star formation they have detected.
Have these surveys detected the majority of
high-redshift star formation, or only
the relatively small fraction that
comparison to the local universe might make us expect?
Although we will be primarily concerned
with the possibility that dust obscuration has hidden
large amounts of star formation from existing surveys,
other physical effects---surface brightness dimming
(e.g. Lanzetta \et 1999) is an obvious example---can
make high-redshift star formation undetectable as well.
The results in this section will provide a rough limit
on the amount of high-redshift star formation
that is be undetected in existing surveys.

This limit can be estimated because
the brightness of the $850\mu$m background provides an upper limit 
on the total amount of star formation at high redshift.  Since
$K$-corrections at $850\mu$m largely cancel out changes in luminosity distance,
a galaxy with given bolometric dust luminosity
will appear almost equally bright at $850\mu$m for any redshift
in the range $1\simlt z\simlt 5$ (e.g. Blain \& Longair 1993).  Taken together with
the observation that star-forming galaxies emit the majority
of their luminosities in the far-IR (cf. Figures 11a--c), this implies
that a flux-limited $850\mu$m survey
is nearly equivalent to star-formation limited survey at $1\simlt z\simlt 5$,
and that the integrated $850\mu$m background provides upper limit to
the total amount of star formation at high redshift.  It is an upper limit,
rather than a measurement,
because some fraction of the $850\mu$m background
is known to be produced by objects other than star-forming
high-redshift galaxies (e.g., AGN and low-redshift galaxies: Edge \et 1999; Ivison \et 2000).

Our strategy for placing limits on the amount of undetected
star formation at high redshift will be to calculate
the contribution to the $850\mu$m background from
the various known high-redshift populations represented in Figures 11b and 11c,
and see if there is significant shortfall between the
background contribution from these galaxies and
the total background that is observed.

\subsection{Sub-mm sources}
We begin by reviewing the contribution to the $850\mu$m background
from resolved sub-mm sources.  According to Barger, Cowie, \& Sanders (1999),
20--30\% of the $850\mu$m background is produced by objects
brighter than SCUBA's 2mJy confusion limit.  Understanding the
nature of these objects is difficult, because the large
diffraction disk of SCUBA on the JCMT means that most
have several plausible optical counterparts,
but multi-wavelength observational programs (e.g. Frayer \et 1999,
Dey \et 1999, BCR, Chapman \et 2000, Frayer \et 2000, Ivison \et 2000) have
established robust optical counterparts for a handful
of them.  This work has shown that approximately
half of the detected $850\mu$m sources are associated 
with AGN (Ivison \et 2000).  Of the sources
which do not appear to be AGN, three---SMMJ14011+0252
(Frayer \et 1999), HR10 (Dey \et 1999), and west MMD11 (Chapman \et 2000)---are
robustly associated with high-redshift star-forming galaxies.
These galaxies, shown in Figures 11b and 11c, have relatively
large dust obscurations $30<L_{\rm bol,dust}/L_{\rm UV}<1000$
and large implied star-formation rates, but their UV luminosities
are comparable to those of typical optically selected galaxies at similar redshifts.
BCR have shown that the brightest $850\mu$m sources tend
to be even more heavily obscured:  their analysis suggests
that $\sim$ 75\% (6/8) of $850\mu$m sources brighter than 6mJy have
extreme dust obscurations $300<L_{\rm bol,dust}/L_{\rm UV}<3000$,
(cf. Figures 11b and 11c).  The ratio of dust to UV luminosity
for these sources is very uncertain because their redshifts
(and consequently luminosity distances) are unknown, but BCR's
radio/sub-mm photometric redshifts (cf. \S 2.4)
would have to be seriously in error to bring their estimated
dust obscurations into a less extreme range.  Despite
the large star-formation rates implied by their sub-mm fluxes,
objects similar to those in the sample of BCR account
for only a tiny fraction, $\sim$ 5\% (Barger, Cowie, \& Sanders 1999),
of the measured $850\mu$m background (Fixsen \et 1998).

The large dust obscurations observed in $850\mu$m sources 
have excited the interest of numerous authors (e.g. Smail \et 1998;
Hughes \et 1998; Eales \et 1999; Barger, Cowie, \& Sanders 1999;
Sanders 1999; BCR), who have rightly noted that if fainter
high-redshift galaxies had similar levels of
dust obscuration, the majority of high-redshift star formation
would have occurred in objects that could not be detected
in any existing survey (cf. Figures 11b and 11c).
This popular argument for ``hidden'' star formation
at high redshift is based on analysis of galaxies responsible for only a small
fraction of the $850\mu$m background, however,
and it ignores the strong correlation of dust obscuration
and luminosity that is observed at low and high redshift
alike (cf. Figure 11).  Compelling arguments
for or against the existence of hidden high-redshift
star formation will require observations of galaxies with
typical star-formation rates, not merely those
with the extreme star-formation rates characteristic
of the objects discussed in this subsection.  Although
some progress can be made at $850\mu$m through observations
of lensed sources (e.g. Smail \et 1998; Ivison \et 2000), the best
constraints currently come from observations at other wavelengths.

\subsection{Mid-IR sources}
At $z\sim 1$, $15\mu{\rm m}$ observations can provide a deeper view of dusty
star formation than the sub-mm observations we have just discussed (cf. Figure 11b).  
The deepest existing large-area $15\mu$m image, 
a $\sim$ 20 square arcmin region centered
on the Hubble Deep Field, was obtained by the ISO satellite in 1997.
In this subsection we will review the properties of the
dusty galaxies at $z\sim 1$ detected in the PRETI reduction
(Aussel \et 1999) of this data.

The primary (high significance) catalog of Aussel \et (1999)
contains 46 $15\mu$m sources.  As discussed by
Aussel \et (1999) and Cohen \et (2000), the majority
of these sources ($\simgt$ 80\%) can be reasonably associated
with relatively bright optical counterparts ($R\simlt 23$).  
Although some of the ISO sources may lie close to
optically bright galaxies by chance, the fact that so many
lie close to optically bright galaxies cannot be
a coincidence.  Spectroscopic redshifts
for 40 of these optical counterparts are provided by Aussel \et (1999)
and Cohen \et (2000); an additional redshift, $z\simeq 0.94$
for HDF-PM3-22, was obtained by Adelberger \et (2000).
16 of the optical counterparts\footnote{HDF-PM3-1, HDF-PM3-6, 
HDF-PM3-8, HDF-PM3-9, HDF-PM3-10, HDF-PM3-11, HDF-PM3-19, HDF-PM3-20, 
HDF-PM3-21, HDF-PM3-22, HDF-PM3-25, HDF-PM3-31, HDF-PM3-34, HDF-PM3-37, 
HDF-PM3-39 and HDF-PM3-45} lie at redshifts $0.8<z<1.2$.

These 16 galaxies provide a reasonably complete sample
of the $z\sim 1$ galaxies in this part of the sky
that have the largest dust luminosities.  The limiting
dust luminosity for this sample is roughly a factor
of three lower than the limit of blank-field SCUBA
surveys.  The sample is not exactly dust-luminosity limited,
because of the scatter in the $L_{MIR}/L_{\rm bol,dust}$
relationship (\S 2.2), and it is not 100\% complete,
because redshifts have not been obtained for some
of the $15\mu$m sources in this field\footnote{If the
$15\mu$m sources without redshifts have the same redshift
distribution as those with redshifts, then perhaps
2 additional $15\mu$m sources in this field lie
at the redshifts of interest $0.8<z<1.2$.}, but
it is as close an approximation a complete,
dust-luminosity sample at high redshift as currently exists.

The bolometric luminosities and dust obscurations of the galaxies
in this sample are indicated in Figure 11b.  Their
typical dust obscurations,
$L_{\rm bol,dust}/L_{UV} \sim 100$, are lower than
those of the brighter sub-mm sources at similar redshifts.
This trend of lower dust obscuration in fainter sources
is also observed in the local universe (Figure 11a).

For the most part, the $z\sim 1$ ISO sources in the HDF appear
to be drawn from among the same population of $z\sim 1$ galaxies
that is routinely studied in the optical.  Most of the sources
are included in the (optical) magnitude-limited HDF
survey of Cohen \et (2000), for example, and 15 of 16
satisfy the Balmer-break photometric selection criteria of Adelberger \et (2000)
and are included in that $z\sim 1$ optical survey as well.
They are not typical of the galaxies in optical populations, as
the non-detection at $15\mu$m of most optically selected
HDF galaxies shows, but Figure 11b suggests
that the $z\sim 1$ ISO population can be naturally interpreted
as the high luminosity, high obscuration tail of
known $z\sim 1$ optical populations.  Optically selected
galaxies are thought to possess a wide range of dust obscurations,
and the analysis of \S 4.3.1 (below) suggests that
$10^{+7}_{-4}$\% have the obscurations $L_{\rm bol,dust}/L_{UV}\simgt 30$
characteristic of ISO sources in the HDF.  For
$\Omega_M=0.3$, $\Omega_\Lambda=0.7$ the comoving
number density of Balmer-break galaxies to $U_n(AB)\sim 25.5$
(roughly the faintest $U_n$ magnitude for HDF ISO sources)
is $\sim 3\times 10^{-2} h^3$ Mpc$^{-3}$, and so
optical observations would lead us to expect that
$\sim 3.0^{+2.0}_{-1.2} \times 10^{-3}$ galaxies
per comoving $h^{-3}$ Mpc$^3$ would have 
dust obscurations $L_{\rm bol,dust}/L_{UV}\simgt 30$
similar to those observed among ISO sources.  This
is roughly equal to the observed comoving number density
of $z\sim 1$ ISO sources in the HDF (16 galaxies with
$0.8<z<1.2$ in $\sim 20$ square arcmin, or
$2.4\times 10^{-3} h^3$ Mpc$^{-3}$ for $\Omega_M=0.3$, $\Omega_\Lambda=0.7$),
showing that the $z\sim 1$ ISO population can be consistently interpreted
as the dustiest part of the optical population.

A worrisome aspect of Figure 11b is that 10 of the
$z\sim 1$ ISO sources are predicted to have detectable
radio fluxes $f_\nu(20{\rm cm})\simgt 50\mu{\rm Jy}$,
while only 2 were detected at this level by Richards (2000).
Other $z\sim 1$ ISO sources are detected in Richards'
20cm image at lower flux levels, but the total
observed 20cm flux for the $z\sim 1$ ISO sources,
after excluding HDF-PM3-6 and HDF-PM3-20 which
have optical spectra suggesting the presence
of an AGN (Cohen \et 2000), is $\sim 360\mu$Jy,
roughly a factor of 4 lower than the total
predicted from the local correlations described
in \S 2.  The implication is that the bolometric
dust luminosities of the ISO sources may be on average
a factor of $\sim 4$ lower than indicated in 
Figure 11b, presumably because the ratio of
mid-IR luminosity to total dust luminosity is
systematically higher at $z\sim 1$ than
in the local universe.  This possibility
only strengthens the main point
we aimed to make in this section, that 
ISO sources have lower dust luminosities $L_{\rm bol,dust}$
and lower dust obscurations $L_{\rm bol,dust}/L_{UV}$
than the brighter SCUBA sources, 
and does not significantly affect the mid-IR/far-UV
comparison of \S 3.3.1 which exploited the
ranks of observed $15\mu$m fluxes rather than their
absolute values; but it does suggest that
estimates of the $850\mu$m background contribution
of ISO sources will be subject to significant uncertainty.

The $850\mu$m background contribution of galaxies
similar to ISO sources can be crudely estimated
as follows.  If we adopt a version of
the mid-IR/$L_{\rm bol,dust}$ relationship (\S 2.2) that
correctly predicts the radio fluxes of $z\sim 1$ ISO sources
in the HDF, and use the sub-mm/$L_{\rm bol,dust}$
relationship of \S 2.1, then the total predicted $850\mu$m flux
from the 16 $z\sim 1$ ISO sources in the HDF is
$\sim 8$mJy, or $0.4$mJy / arcmin$^2$, roughly
3\% of the background measured by Fixsen \et (1998).
This number applies only to galaxies in the
redshift range $0.8<z<1.2$, where ISO $15\mu$m observations
are particularly sensitive due to the PAH features.
Because little is known about the number density
evolution of bright and highly obscured galaxies
similar to ISO sources, it is difficult to estimate
the contribution from galaxies outside this redshift
interval, but a similar population of galaxies distributed uniformly
between $z=1$ and $z=5$ would contribute $\sim 40$\%
($\Omega_M=0.3$, $\Omega_\Lambda=0.7$) of the 
total $850\mu$m background.

Because of the substantial overlap between 
ISO sources and galaxies in optical surveys,
much of the $850\mu$m background contribution 
attributed above to ISO sources will be included as well
in our calculation of the background contribution
from optically selected galaxies, which we now describe.

\subsection{Optical sources}
Relatively few high-redshift galaxies have been
detected by their dust emission, and the few
that have apparently account for less than
half of the $850\mu$m background.  
The vast majority of known high-redshift galaxies have
been found in optical surveys.  Can the large numbers
of optically selected galaxies account for the
remainder of the $850\mu$m background?

Figures 11b and 11c illustrate the relationship
between optically selected galaxies and
the $850\mu$m and $15\mu$m sources we have just discussed.
The solid circles in Figure 11b represent $z\sim 1$ galaxies
from the spectroscopic Balmer-break sample of Adelberger \et (2000).
Except in cases where their dust emission has been directly
measured, the handful of Balmer-break galaxies with estimated
dust obscurations $L_{\rm bol,dust}/L_{UV}> 100$ were omitted
from this plot due to the large uncertainties
in their dust luminosities (cf. \S 3.3).  Balmer-break
galaxies with $U_n(AB)>25.5$ were also omitted, because
at these faint magnitudes the Balmer-break sample
suffers from severe selection effects.  The solid circles
in Figure 11c represent Lyman-break galaxies from the
spectroscopic sample of Steidel \et (2000).  
As described in the introduction to \S 4, the dust luminosities
of the optically selected galaxies in Figures 11b and 11c
were estimated from their dust emission in the few
cases where it was measured, and from their UV photometry
in the typical cases where it was not.

These figures support the common view (e.g. Eales \et 1999; BCR;
Frayer \et 2000) that galaxies detected in the UV
are typically far less luminous and less obscured than
those detected in the IR.  In the next sections
we will see if UV-selected populations can make up
in number what they lack in luminosity and still
produce a significant contribution to the $850\mu$m background.

\subsubsection{The $850\mu$m luminosity function of Lyman-break galaxies}
As a first step we will estimate the contribution to the $850\mu$m background
from the best understood population of UV-selected high-redshift
galaxies:  the Lyman-break galaxies at $z\sim 3$.
Figure 12 shows these galaxies' apparent magnitude
and $\beta$ distributions (Steidel \et 1999, Adelberger \et 2000).
Adelberger \et (2000) find no significant correlation between
apparent magnitude and $\beta$ for Lyman-break galaxies
in their ground-based sample (${\cal R}\leq 25.5$; a scatter
plot of UV-luminosity versus dust obscuration in $z\sim 3$ galaxies is presented
in Figure 17 and discussed in \S 4.4 below)
and so it is relatively simple to turn these two distributions
into an $850\mu$m luminosity function.  The result is
shown in Figure 13.  To produce this plot in a way
that accounted for the various uncertainties, we first
generated a large number of random realizations of Lyman-break
galaxies' apparent magnitude and $\beta$ distributions, consistent
to within the errors with the distributions of Figure 12.
We then picked at random a $\beta$ distribution and an
apparent magnitude distribution from among these realizations,
generated a long list of ${\cal R}$, $\beta$ pairs, with
${\cal R}$ drawn randomly from the best Schechter-function fit to
the apparent magnitude distribution and $\beta$ from the $\beta$ distribution,
assigned to each ${\cal R}$, $\beta$ pair in the list
an $850\mu$m flux with eq. 8,
and adjusted each of these $850\mu$m fluxes
at random by an amount reflecting the uncertainty
in their predicted values (see \S 3).  Binning this list 
and dividing by the appropriate volume produced
one realization of the $850\mu$m luminosity function shown in Figure 13.
We repeated the process, choosing a different realization of the
$\beta$ and apparent magnitude distributions each time, until we had
many binned realizations of the expected $850\mu$m luminosity function.
The points shown in Figure 13 are the mean values among
these many binned realizations; the uncertainties are the standard deviations.
Also shown in Figure 13 are two related luminosity functions estimated
in a similar manner:  the
expected $60\mu$m luminosity function of Lyman-break galaxies (similar
in shape to their bolometric dust luminosity function) and
the ``dust corrected'' UV luminosity function (derived
by taking $A_{1600} = 4.43+1.99\beta$ for the extinction
in magnitudes at 1600\AA, as suggested by MHC).  The shapes
of these two luminosity distributions, both essentially
equivalent to the star-formation rate distribution,
are similar to what is predicted by the simplest
hierarchical models of galaxy formation (Adelberger 2000, Adelberger \et 2000).

The systematic uncertainties in these luminosity functions
are far larger than the random uncertainties.
The most obvious systematic uncertainty is in the validity of our underlying assumption
that the $\beta$/far-IR correlation will hold at high redshift.
This uncertainty is difficult to treat quantitatively.  If the $\beta$/far-IR correlation
does not hold these results are meaningless; the data in \S 3 give us hope that
it may hold; there is little else to say.

A second source of systematic
uncertainty is the unknown shape of the apparent magnitude distribution
at magnitudes fainter than ${\cal R}\sim 27$.  Galaxies so
faint in the UV may still have large star-formation rates
if they are sufficiently dusty, and so our estimate of the
number density of high star-formation rate galaxies depends to some
degree on the unknown number density of these faint galaxies.
In our calculation above we assumed that the $\alpha\sim -1.6$
faint-end slope of the apparent magnitude distribution would continue
to arbitrarily faint magnitudes.  This extrapolation has a significant
effect on our conclusions only at the faintest luminosities;
Figure 13 shows the luminosities below which most of
the estimated luminosity density comes from objects
with ${\cal R}>27$.  These parts of the luminosity
functions should be viewed with considerable scepticism.

A related source of systematic uncertainty is our assumption
that the $\beta$ distribution of Lyman-break galaxies
is the same at all apparent magnitudes.  This is known
to be true only for ${\cal R}<25.5$ (Adelberger \et 2000),
and so the faintest points in our inferred luminosity
functions are subject to further systematic uncertainties
which are difficult to quantify.  This source of systematic
uncertainty is negligible only at the brightest luminosities,
where most of the estimated luminosity density comes from
galaxies with ${\cal R}<25.5$; see Figure 13.

The final source of significant systematic uncertainty
is the shape of the $\beta$ distribution.  Deriving this
distribution from our data is difficult.
It requires (among other things) a quantitative
understanding of the selection biases that result
from our photometric selection criteria and
of the effects of photometric errors and Lyman-$\alpha$ emission/absorption
on our measured broadband colors.  Our current best estimate
of the $\beta$ distribution is shown in Figure 12 (Adelberger \et 2000).  To give
some idea of the systematic uncertainty in this distribution,
we also show our previously published estimate 
(adapted from Steidel \et 1999) and the $\beta$ distribution
we would derive if we neglected photometric errors and
incompleteness corrections altogether and simply
corrected the measured broadband colors for the
spectroscopically observed Lyman-$\alpha$
equivalent width.  The most important difference
between these three distributions (for our present
purposes) is the shape of their red tails. The true
shape almost certainly falls within the range
they span; it is unlikely to be redder than
the distribution of Steidel \et (1999), and, since 
all known selection biases cause the Lyman-break technique
to miss the reddest objects at $z\sim 3$,
it cannot be bluer than the distribution that neglects
completeness corrections.
The luminosity functions
that we would infer given these two limiting $\beta$ distributions
are show in Figure 13 by curved lines.  The systematic
uncertainty in the $\beta$ distribution means that
the mean extinction at 1600\AA\ for Lyman-break galaxies,
a factor of 6 in our best estimate, could lie between
a factor of 5 and a factor of 9.

\subsubsection{UV-selected populations and the $850\mu$m background}

In this section we will extend the preceding calculation to derive
a crude estimate of the total contribution to the $850\mu$m background
from known UV-selected populations at $1<z<5$.
\S 4.3.1 described the uncertainties affecting
our calculation of $z\sim 3$ Lyman-break galaxies' contribution
to the $850\mu$m background. Because relatively little is known about
UV-selected populations at redshifts other than $z\sim 3$,  
the uncertainties in
the present calculation will be much larger still.
As a first approximation we
can assume that these galaxies will have
the same $\sim 1800{\rm\AA}$ luminosity distribution,
the same $\beta$ distribution, and the same
lack of correlation between $\beta$ and $\sim 1800{\rm\AA}$
luminosity as Lyman-break galaxies at $z\sim 3$.
This is clearly an over-simplification, but
the available data suggest that it might not be
seriously incorrect.
For example, Lyman-break galaxies at $z\sim 4$ appear to have
approximately the same luminosity distribution and (possibly)
$\beta$ distribution as Lyman-break galaxies at
$z\sim 3$ (Steidel \et 1999); Balmer-break galaxies
at $z\sim 1$ have far-UV spectral slopes $\beta$ that
lie mainly in the range $-2.2<\beta<-0.5$ observed
in Lyman-break galaxies at $z\sim 3$ (Adelberger \et 2000);
and, like Lyman-break galaxies at $z\sim 3$,
Balmer-break galaxies at $z\sim 1$ exhibit no significant correlation of 
$\beta$ with apparent magnitude
at rest-frame $\sim 1800{\rm \AA}$ (Adelberger \et 2000;
cf. Figure 17, which we discuss below).
The available data suggest further that the comoving
star-formation density in UV-selected populations
is roughly constant for $1<z<5$ (Steidel \et 1999),
and so the contribution
to the $850\mu$m background from UV-selected high-redshift populations
can probably be crudely approximated as
the contribution that would
arise from Lyman-break--like populations distributed
with a constant comoving number density at $1<z<5$.

To estimate this contribution we simulated
an ensemble of $\sim 10^6$ Lyman-break--like
galaxies at $1<z<5$.  Each galaxy was assigned
a UV slope $\beta$ drawn randomly
from the (shaded) $\beta$ distribution shown in Figure 12;
a redshift $z_i$ drawn randomly from the interval
$1<z<5$ with probability proportional to $dV/dz$
(where $dV/dz$ is the comoving volume per arcmin$^2$
per unit redshift in an $\Omega_M=0.3$, $\Omega_\Lambda=0.7$
cosmology); an apparent magnitude at rest-frame $1600{\rm\AA}$,
$m_{1600}$, drawn randomly from the apparent magnitude
distribution in Figure 12 shifted by the
($\Omega_M=0.3$, $\Omega_\Lambda=0.7$) distance
modulus between $z_i$ and $z=3.0$ and truncated at $m_{1600}=27$; and
an observed-frame $850\mu$m flux with a version of eq. 8 appropriate
to redshift $z_i$.  Finally each $850\mu$m flux was adjusted at random
by an amount reflecting the uncertainty in its predicted
value (\S 3).  

Figure 14 was produced by placing these simulated galaxies
into bins of $850\mu$m flux and dividing by the
appropriate area.  The top panel shows the estimated
contribution to the $850\mu$m background per logarithmic
interval in $f_\nu$, $f_\nu^2 n(f_\nu)$, where
$n(f_\nu)$ is the number of sources per mJy per
square degree.  The overall background expected
from UV-selected populations in this crude calculation,
$4.1\times 10^4 {\rm mJy / deg}^2$, is 
close to the measured background of $4.4\times 10^4$
(Fixsen \et 1998).  The closeness of the agreement
is not especially meaningful, because the uncertainties
in the calculation are so large, but nevertheless it is clear
that the known UV-selected galaxies
at high redshift could easily have produced
the $\sim$ 75\% of the $850\mu$m background
that has not currently been resolved into discrete sources
by SCUBA.  There is little need to invoke
large amounts of hidden high-redshift star formation
in order to account for the brightness of
the $850\mu$m background.  

If the bulk of the $850\mu$m background is produced
by galaxies similar to those in UV-selected surveys,
as we have argued, then the shape of the $850\mu$m
number counts should be similar to what our calculation suggests
UV selected populations would produce.
This appears to be the case.  The bottom panel
of Figure 14 compares the cumulative $850\mu$m number counts
$N(>f_\nu)$ ($\equiv$ the number of objects per square degree
brighter than $f_\nu$) measured by Blain \et (1999b)
to the number counts that would be produced by the UV-selected high-redshift 
galaxies in our calculation.  Although there are clear
differences at the brightest end, where the observed
number counts receive contributions from AGN, from low-redshift galaxies, and
from extremely luminous high-redshift galaxies that may differ
from the more normal galaxies detected in the UV,
the observed number counts are remarkably close
to what UV-selected populations would produce
in the range $0.3<f_\nu<3$mJy that dominates
the $850\mu$m background.

Figure 15 shows the expected contribution to the $850\mu$m 
background from objects with different apparent
magnitudes at rest-frame 1600\AA.  This gives a rough
idea of the apparent magnitudes we would expect to measure
for sub-mm sources with various fluxes if rest-frame UV-selected
populations obeying the local correlations between far-UV and
far-IR fluxes produced the bulk of the $850\mu$m
background.  At any given
850$\mu$m flux density, objects are expected to have a wide range of 
redshifts and dust obscurations, and consequently of
optical magnitudes, but even at the brightest $850\mu$m fluxes
most objects are expected to be relatively faint ($m_{1600}>24$)
in the optical.  The observed faintness of optical counterparts
to $850\mu$m sources is sometimes thought to show that the
$850\mu$m background is produced by objects not present
in rest-UV-selected surveys, but this argument has to be made carefully:
UV-selected high-redshift galaxies are themselves
quite faint in the optical, and as a result it is not sufficient to
show (for example) that most optical counterparts are
fainter than $24^{\rm th}$ magnitude; it must be shown
instead that they are significantly fainter than Figure 15
predicts.  The available data suggest that the sources
responsible for brightest $850\mu$m number counts,
$f_\nu\simgt 6$mJy, are in fact significantly fainter than
the predictions of Figure 15 (e.g. BCR; Frayer \et 2000), 
and this is hardly surprising since our calculation
shows that UV-selected populations cannot by themselves
account for the brightest $850\mu$m number counts.
Objects with larger obscurations $L_{\rm bol,dust}/L_{UV}\sim 1000$
are required and have been found.  Measuring the
optical magnitudes of $\sim 1$mJy sources would provide a better test of
whether UV-selected populations contribute significantly to 
the far-IR background, but with current technology this is possible
only for lensed sources and adequate data do not yet exist.

Also shown in Figure 15 is the contribution to the $850\mu$m background
that would be produced by galaxies with $27<m_{1600}<29$ if
the $\alpha=-1.6$ faint-end slope of the luminosity
function continued past the faintest observed magnitude of
$m_{1600}\simeq 27$ and if dust obscuration and
reddening remained uncorrelated at the faintest magnitudes.  
Due to the steep faint-end slope, objects with
$m_{1600}>27$ might be expected to contribute significantly
to the $850\mu$m background.  MHC have presented
evidence, however, from Lyman-break galaxies with $25.5\simlt m_{1600} < 27$
in the HDF, that galaxies fainter in the UV tend to have lower dust
extinctions.  
Because apparent magnitude and dust obscuration
are not correlated among $z\sim 3$ Lyman-break galaxies
with $m_{1600}<25.5$, we assumed when producing
Figures 14 and 15 that they would remain uncorrelated
at fainter magnitudes as well.  MHC's analysis
suggests that this assumption is incorrect:
galaxies with $25.5<m_{1600}<27$ are probably
somewhat less dusty than we have assumed and galaxies
with $m_{1600}>27$ are probably significantly less dusty.
When estimating the total background contribution
from UV-selected populations (Figure 14),
we assumed as a compromise that
galaxies with $25.5<m_{1600}<27$ would
be as dusty as galaxies with $m_{1600}<25.5$, and that
galaxies with $m_{1600}>27$ would contain negligible
amounts of dust.  This is obviously a very crude
solution to a complicated problem; the background contribution from objects
with $m_{1600}>25.5$ is a
significant source of systematic uncertainty in Figures 14 and 15.

Figure 15 would be more useful for observational tests if we
showed the predicted magnitudes at fixed observed-frame
wavelengths, say the $i$ band, rather than a fixed rest-frame
wavelength, but this requires us to estimate $K$-corrections
for these galaxies and we have not found an adequate model
for their UV/optical SEDs.  The problem is illustrated
by Figure 16, which shows the measured rest-frame UV/optical SEDs
of two Lyman-break galaxies detected at $850\mu$m,
SMMJ14011+0252 (Smail \et 1998) and west MMD11 (Chapman \et 2000). 
The optical data were obtained as described in \S 3.1
and in Steidel \et (2000); the near-IR photometry
was obtained with NIRC (Matthews \& Soifer 1994) on Keck I
with excellent seeing ($\simlt 0\secpoint 5$) in May 1999.
Neither SED is fit well by a model of
a continuously star-forming galaxy subjected to
varying amounts of dust following a standard extinction law.
Moreover, even though SMMJ14011+0252 and west MMD11 have
essentially indistinguishable SEDs in the UV---both
are consistent with the same value of $\beta$---their
optical SEDs are significantly different.
The shapes of galaxies' UV/optical SEDs are strongly
dependent on their star-formation histories, and perhaps
the simplest conclusion to draw from Figure 16
is that the star-formation histories of high-redshift
galaxies are complicated and diverse.
We have found no obvious way to predict their
optical photometry from their UV photometry.  See
Shapley \et (2000) for a more complete discussion.

This is a good place to remind readers that our
UV-based estimates of high-redshift galaxies' bolometric
dust luminosities are not derived from an
assumed intrinsic SED and dust reddening law; they are derived 
instead from the empirical (and somewhat mysterious)
correlation between $\beta$ and $L_{\rm bol, dust}/L_{1600}$
described in \S 2.3.   The diversity of observed
SEDs makes it difficult to understand why this
correlation should exist, but local (\S 2)
and (less convincingly) high redshift (\S 3) observations suggest that
it does.  

\subsection{Discussion}
The major result of this section is our calculation
showing that galaxies similar to those
detected in UV-selected surveys probably produced
a large fraction of the measured $850\mu$m background.
Recent analyses of $850\mu$m surveys
have reached the opposite conclusion
(e.g. Smail \et 1998, Hughes \et 1998, Eales \et 1999, Barger, Cowie, \& Sanders 1999,
Sanders 1999, BCR).  
Skeptical readers may wonder if our calculation
is robust given its apparently heavy reliance
on the poorly tested assumption (\S 3) that high-redshift
galaxies' dust luminosities can be estimated
from their UV fluxes with the correlations of \S 2.

Although the details of our calculation certainly
depend upon the assumption that
high-redshift galaxies obey MHC's $\beta$/far-IR
relationship, the overall background contribution
from UV-selected galaxies depends primarily
upon the adopted mean value of $L_{\rm bol,dust}/L_{UV}$.
MHC's relationship implies that
$\langle L_{\rm bol,dust}/L_{UV} \rangle \simeq 1.66\times\langle 10^{0.4A_{1600}}-1\rangle \sim 8$
for galaxies at $z\sim 3$, and our calculation in \S 4.3
showed that this factor of 8 is sufficient
for UV-selected galaxies to have produced the bulk
of the $850\mu$m background.
High-redshift galaxies may not satisfy MHC's relationship,
but the adopted $\langle L_{\rm bol,dust}/L_{UV} \rangle \sim 8$
is still a plausible estimate of their mean dust obscuration.
This mean obscuration
is hardly unrealistically high; it is roughly equal to
mean obscuration among the Lyman-break galaxies observed
in the sub-mm by Chapman \et (2000), for example,
and it is lower than the mean obscuration ($\sim 15$)
in the local UV-selected starburst sample of MHC.  Even the local
spirals in the sample discussed before \S 4.1 have
$\langle L_{\rm bol,dust}/L_{UV}\rangle\sim 5$; given the
correlation of dust obscuration and star-formation rate
that appears to exist at all redshifts (cf. Figures 11a--c),
it is hard to imagine that a rapidly star-forming population
at any redshift could have values of $L_{\rm bol,dust}/L_{UV}$
significantly lower than spirals in the local universe.

But the fundamental difference between our analysis
and the analyses of previous authors who
reached opposite conclusions is not
simply the value of $\langle L_{\rm bol,dust}/L_{UV}\rangle$ adopted
for UV-selected galaxies; it is instead the assumed
dependence of dust obscuration on luminosity among
high-redshift galaxies.  Arguments for large
amounts of hidden star-formation at high redshift
generally assume that dust obscuration and
luminosity are independent.  Data from low
and high redshifts lead us to believe
that $L_{\rm bol,dust}/L_{UV}$\lower.5ex\hbox{$\;\buildrel\propto\over\sim\;$}$L_{\rm bol,dust}$
is a better approximation (cf. Figures 11a--c and 17).
This correlation between dust obscuration and luminosity
means that the $f_\nu(850\mu{\rm m})\sim 1$mJy sources
that produce most of the $850\mu$m background
are likely to be significantly less obscured than
the brighter sources currently detectable in the sub-mm.
For example, if the $\sim 10$mJy sources studied
by BCR have $200\simlt L_{\rm bol,dust}/L_{UV}\simlt 2000$
(cf. Figures 11b,c), then the correlation of dust obscuration
and luminosity would suggest the 1mJy sources responsible
for most of the $850\mu$m background have 
$20\simlt L_{\rm bol,dust}/L_{UV}\simlt 200$.  These
sources would be dustier than typical optically selected galaxies,
but they would nevertheless be as easy to detect 
in the optical as in the sub-mm.
More generally, the correlation of dust obscuration
with luminosity means that heavily obscured sources
are not necessarily any harder to detect
in the rest-frame UV than less obscured sources.  Their
higher UV obscurations are largely cancelled out
by their higher intrinsic luminosities.  This is illustrated
by Figure 17, which shows the observed UV luminosities
of sources with different dust obscurations
at low and high redshift.  Observed UV luminosities
are similar for galaxies with dust obscurations $L_{\rm bol,dust}/L_{UV}$
spanning four orders of magnitude.

How robust is the claimed existence of the correlation between
dust obscuration and luminosity?  At low redshift,
where objects' dust luminosities are well constrained
by IRAS measurements, its existence is indisputable.
At higher redshifts, dust luminosities can only
be estimated with the uncertain empirical correlations
described in \S 2, and its existence is somewhat less
secure.  But even if none of the empirical correlations
of \S 2 hold at high redshift, it is difficult to 
escape the conclusion that dust obscuration and luminosity
are correlated.  Without the correlations of \S 2
we would not know precisely where on the $x$-axis
to place any object in Figure 17, for example,
but we would still know that ISO sources are typically more
luminous and more obscured than optical sources,
and that SCUBA sources are typically more
luminous and more obscured than ISO sources\footnote{
For example, typical optical sources at $z\sim 1$ are fainter at
$15\mu$m and 20cm than ISO sources at $z\sim 1$; typical $z\sim 1$
ISO sources must be fainter than typical SCUBA sources,
or else they would over-produce the $850\mu$m number counts.}; 
and that is enough to establish the existence of
the correlation.

A final and more direct argument that
UV-selected galaxies contribute significantly
to the $850\mu$m background comes from the work of
Peacock \et (2000), who showed
that positive background fluctuations in a deep $850\mu$m
image of the HDF (Hughes \et 1998) tend to
be coincident with the locations of UV-selected high-redshift
galaxies.  Their detailed analysis suggests that UV-selected
populations must have produced at least one quarter, and probably
at least one half, of the total $850\mu$m background.

Taken together the preceding arguments make a reasonably
strong case that the majority of high-redshift star formation
is detectable in the deepest optical surveys.  But
this does not imply that IR/sub-mm observations are unnecessary.
Although many of the results that have emerged from sub-mm surveys
could have been inferred from UV data alone---the brightness
of the $850\mu$m background, for instance, or the domination 
of the $850\mu$m background 
by $\sim 1$mJy sources (e.g. Blain \et 1999b; cf. Figure 14), 
or the $\sim$ thousand-fold increase in number density
of ULIRGs from $z\sim 0$ to $z\simgt 3$ 
(e.g. Sanders 1999; cf. Figure 13)---some could not.
For example, there was little indication from UV-selected
surveys that extremely dusty galaxies with
$L_{\rm bol,dust}/L_{UV}\gg 100$ even existed (cf. Figure 12), let
alone that they hosted significant fraction ($\simgt 5$\%)
of high-redshift star formation.
Observations in the IR will continue
to play a central role in studies of high-redshift
star formation, both as tests of the dust corrections
that UV-selected surveys require and as the most
efficient way to find the dustiest and most rapidly
star-forming galaxies.

\section{SUMMARY}

The available evidence strongly suggests that dust in high-redshift
galaxies will have absorbed most of the UV radiation
emitted by their massive stars.  Dust certainly absorbs
most of the UV radiation emitted by rapidly star-forming
galaxies in the local universe, and the ratio of
the far-IR to optical backgrounds shows that the
situation must have been similar throughout the history
of the universe.  Estimating the bolometric dust
luminosities of high-redshift galaxies is therefore
a crucial step in deriving their star-formation rates.
The first part of this paper was primarily concerned with ways 
bolometric dust luminosities can be estimated even
though most of the dust emission is blocked by the earth's atmosphere
before it can be detected on the ground.

We began in \S 2 by describing correlations between
local galaxies' bolometric dust luminosities and
luminosities at more accessible far-UV, mid-IR, sub-mm, and radio
wavelengths.  These correlations provide the basis
for attempts to estimate high-redshift galaxies'
bolometric dust luminosities.  Much of this section
was simply a review of previously published results;
the minor new contribution was a quantification
of the uncertainties involved in estimating bolometric
dust luminosities from observations at a single
rest wavelength $25\mu{\rm m}\simlt\lambda\simlt 600\mu{\rm m}$
or $\lambda\sim 8\mu{\rm m}$.  The data
suggest that radio, sub-mm, mid-IR, and far-UV
estimates of local galaxies' bolometric
dust luminosities are all subject to a comparable uncertainty of 0.2--0.3 dex.

\S 3 was concerned with establishing whether the bolometric
dust luminosities of high-redshift
galaxies can be estimated
from their fluxes through various atmospheric windows
with the correlations of \S 2.  The evidence presented
was of uneven quality.  The galaxy with the
highest quality observations, SMMJ14011+0252 at
$z=2.565$, obeyed the expected correlations nicely.
The observations of more typical $z\sim 3$ galaxies
had extremely low signal to noise ratios ($\simlt 1$),
and were largely inconclusive, though at least
consistent with the expected correlations.
Galaxies in our $z\sim 1$ sample
were bright enough in the mid-IR and radio for us to show
with high significance ($>$99\%) that their predicted and observed fluxes
in each of these bands were positively
correlated.  The data suggested moreover that this positive correlation
was at least broadly consistent with the expected linear correlation
$\langle f_{\rm observed}\rangle =\langle f_{\rm predicted}\rangle$.
Taken together the data in this section make a plausible but still
unconvincing case that the bolometric dust luminosities
of high-redshift galaxies can be estimated with the expected
accuracy using correlations calibrated in the local universe.

Our most important results were presented in the second part of this paper, \S 4,
where we estimated the contributions to the far-IR
background from different high-redshift populations.
The brightness of the $850\mu$m background
provides a rough measure of the total amount
of star formation at high redshift, and
a significant shortfall between the
background contribution from known populations
and the measured background would show
that existing surveys have not detected
the majority of star formation at high redshift.
We did not find any evidence
for a shortfall.  The extremely luminous objects
currently detectable in the sub-mm can account for
$\sim 25$\% of the far-IR background; the somewhat fainter
galaxies detected in the mid-IR by ISO can
account for perhaps an additional $\sim 15$\%; and
the galaxies detected in optical surveys
can account for the remainder.  Attributing the far-IR background
to different galaxy populations in this way
obscures the fact that there is substantial overlap between galaxy populations
selected at different wavelengths, especially the
mid-IR and optical, but the important point is that
the brightness of the $850\mu$m background is similar to
what known high-redshift populations would
produce if they obeyed the local correlations of \S 2.
This casts doubt on recent claims
that the majority of high-redshift star formation
is hidden from existing surveys (e.g. Lanzetta \et 1999; BCR).

The analysis of \S 4 suggested that the bulk
of the $850\mu$m background was produced
by moderately obscured galaxies ($1<L_{\rm bol,dust}/L_{UV}<100$)
similar to those that host most of the star formation
in the local universe and to those that are detected
in UV-selected high-redshift surveys.
The brightness of the $850\mu$m background is
sometimes cited as evidence that most of the
star formation in the universe occurred
in dusty objects missing from UV-selected surveys,
but this is a mistaken conclusion:
the bright $850\mu$m background
shows only that a large fraction of star formation
occurred in dusty objects, not that it occurred
in the extremely dusty objects ($L_{\rm bol,dust}\simgt 100 L_{UV}$)
that are easier to detect in the sub-mm than the far-UV.  
If we assumed that star formation at high redshift
occurred only in known UV-selected populations
and asked how dusty these galaxies would have to be
to produce the observed $850\mu$m background, we would
find that the mean required extinction at 1600\AA\ was
a factor of $\sim 6$---precisely the value that is
generally inferred for UV-selected galaxies at
$z\sim 1$ and $z\sim 3$ and similar to the values observed in 
spirals and UV-selected starbursts
at $z\sim 0$.  The brightness of the $850\mu$m background
does not by itself require any star formation at all
to have occurred in galaxies dustier than the moderately
obscured ones ($1<L_{\rm bol,dust}/L_{UV}<100$) known to
contain most of the star formation in the local universe.

This is not to say that extremely dusty galaxies ($L_{\rm bol,dust}/L_{UV}>100$)
do not exist.  Some clearly do (e.g. Dey \et 1999; BCR;
Frayer \et 2000).  Part of the far-IR background
is produced by them.  Our argument is simply that if
high-redshift galaxies are similar to local galaxies
in either their typical dust obscurations or their correlation
of dust obscuration with luminosity, then UV-selected
populations at high redshift must have contributed
significantly to the $850\mu$m background (see \S 4.5).
The fundamental difference between our analysis and the
analyses of previous authors who reached opposite
conclusions is the assumption made about
the correlation between luminosity and dust obscuration.
We have assumed that a correlation exists; they have
assumed that it does not.  The assumed existence of this
correlation strongly affects the amount of high-redshift
star formation that is detectable in the optical, since
it implies first that the $0.3<f_\nu(850\mu{\rm m})<3$mJy sources which
dominate the $850\mu$m background are likely to be
less obscured than the brighter sources detectable with
SCUBA, and second that the dustiest, most rapidly star-forming galaxies at
high redshift are not necessarily any harder to detect in the UV than
the less obscured sources forming stars at more moderate rates.
A correlation similar to the one we have assumed is known
to exist in the local universe, and in \S 4 we presented
evidence that it exists at $z\sim 1$ and $z\sim 3$ as well.
As illustrated by Figure 17,
the high-redshift galaxies that have been detected
by their dust emission are (with few exceptions) no fainter in the UV
than the less obscured UV-selected galaxies at comparable redshifts.

Although we have argued that
the majority of high-redshift star formation is detected in UV-selected
surveys, and that many of the recent results from far-IR/sub-mm observations
could have been predicted from UV observations alone
(e.g. the brightness of the $850\mu$m background, the shape of the
$850\mu$m number counts, the $\sim$ thousand-fold increase in the
number density of ULIRGs at high redshift), we do not want to
suggest that IR observations are unnecessary.  Our arguments
imply on the contrary that they are indispensable.  Large
corrections for dust extinction will be necessary in the
interpretation of UV-selected surveys, and only IR observations
can show whether the currently adopted corrections are valid
or suggest alternatives if they are not. 
And the fact remains that IR/sub-mm surveys are the most efficient
way to find the extremely luminous and dusty galaxies
known to produce at least 5\% of the $850\mu$m background.

It is commonly assumed that star formation at high redshift
occurs in two distinct populations of galaxies, one relatively
unaffected by dust and the other completely obscured.
We favor a different view.  High-redshift galaxies
exhibit a continuum of dust properties.
Some contain little dust, and some contain so much dust
that they are nearly invisible in the optical; but
the majority of star formation occurs in galaxies between
these two extremes, in galaxies with $1<L_{\rm bol,dust}/L_{UV}<100$.
The galaxies in this dominant population are undeniably dusty---most emit
a larger fraction of their bolometric luminosities
in the far-IR than in the far-UV---but with current technology they
are easiest to detect in the rest-frame UV.
Because in most cases only a small fraction of their
luminosities emerge in the far-UV, large corrections for dust obscuration
are required when estimating their star formation rates.
A major purpose of this paper was to show
that these corrections are tractable and produce sensible results. 
Although the validity of UV-derived dust corrections has
not been convincingly demonstrated on an object-by-object
basis, and further observations in the infrared are
clearly required, our analysis suggests that the majority
of high-redshift star formation occurs in objects
which can at least be detected in existing UV-selected
surveys.
This is good news for those attempting to understand 
galaxy formation at high redshift, since it is likely 
that UV-selected surveys will continue to provide the most 
statistically robust information for at least the next decade.  

\bigskip
\bigskip

This paper was made possible by several researchers who generously
gave us access to their data during the early
stages of its publication.  We are grateful to D. Calzetti
for her ISO measurements of dust emission from local starbursts,
to E. Richards for his 20cm image of the Hubble Deep Field, and to T. Haynes,
G. Cotter, and A. Bunker
for the optical magnitudes of HR10.
In addition M. Dickinson, M. Giavalisco, M. Pettini, A. Shapley,
and R. Brunner helped obtain some of the data we have presented,
and G. Taylor kindly provided us with an estimate of SMMJ14011's
radio flux prior to its publication by other authors.
D. Elbaz recommended looking at the $15\mu$m fluxes
of galaxies in our Balmer-break sample.  D. Frayer and A. Barger are
thanked for several informative conversations.  Useful comments
on an earlier draft were provided by A. Blain, M. Dickinson, D. Elbaz, 
M. Pettini and (especially) S. Lilly, the referee.
CCS acknowledges support from the U.S. National Science Foundation through
grant AST 95-96229, and from the David and Lucile Packard Foundation.
This research has made use of the NASA/IPAC Extragalactic Database (NED)
which is operated by the Jet Propulsion Laboratory, California Institute of Technology,
under contract with the National Aeronautics and Space Administration.

\bigskip
\begin{deluxetable}{cccc}
\tablewidth{0pc}
\scriptsize
\tablecaption{Local Dust SEDs\tablenotemark{a}}
\tablehead{
	\colhead{$\lambda(\mu{\rm m})$} &
	\colhead{$\langle 1 / {\cal K}_\lambda\rangle$} &
	\colhead{$\sigma_{\cal K} ({\rm dex})$} &
	\colhead{$N_{\rm objects}$}
}
\startdata
25 & 0.25 & 0.21 & 27\nl
50 & 0.64 & 0.08 & 27\nl
75 & 0.69 & 0.06 & 27\nl
100 & 0.53 & 0.12 & 27\nl
125 & 0.38 & 0.17 & 27\nl
150 & 0.26 & 0.22 & 27\nl
175 & 0.17 & 0.25 & 27\nl
200 & 0.11 & 0.30 & 27\nl
225 & 0.076 & 0.30 & 21\nl
250 & 0.055 & 0.31 & 21\nl
275 & 0.041 & 0.32 & 21\nl
300 & 0.030 & 0.32 & 21\nl
325 & 0.023 & 0.32 & 21\nl
350 & 0.017 & 0.32 & 21\nl
375 & 0.013 & 0.32 & 21\nl
400 & 0.010 & 0.32 & 21\nl
425 & 0.0080 & 0.33 & 21\nl
450 & 0.0063 & 0.33 & 21\nl
500 & 0.0040 & 0.33 & 21\nl
550 & 0.0026 & 0.33 & 21\nl
600 & 0.0017 & 0.33 & 21\nl
650 & 0.0012 & 0.33 & 21\nl
\enddata
\tablenotetext{a}{These numbers are appropriate only for
rapidly star-forming galaxies similar to those in
the sample of \S 2.1.  The dust SEDs of more
typical star-forming galaxies (such as the Milky Way)
are significantly different.}
\end{deluxetable}

\bigskip
\begin{deluxetable}{lcccccl}
\tablewidth{0pc}
\scriptsize
\tablecaption{Photometry}
\tablehead{
	\colhead{Object} & \colhead{G\tablenotemark{a}} &
	\colhead{{\cal R}\tablenotemark{a}} &
	\colhead{i\tablenotemark{a}} &
	\colhead{J\tablenotemark{b}} &
	\colhead{H\tablenotemark{b}} &
	\colhead{$K_s$\tablenotemark{b}}
}
\startdata
SMMJ14011\tablenotemark{c}& 21.94 & 21.25 & -     & 19.33 & 18.44 & 17.69\nl
West MMD11	          & 25.09 & 24.05 & 23.84 & 21.28 & 20.12 & 19.44\nl
\enddata
\tablenotetext{a}{AB system}
\tablenotetext{b}{Vega system}
\tablenotetext{c}{Sum of the components J1 \& J2}
\end{deluxetable}
\newpage
\begin{figure}
\figurenum{1}
\plotone{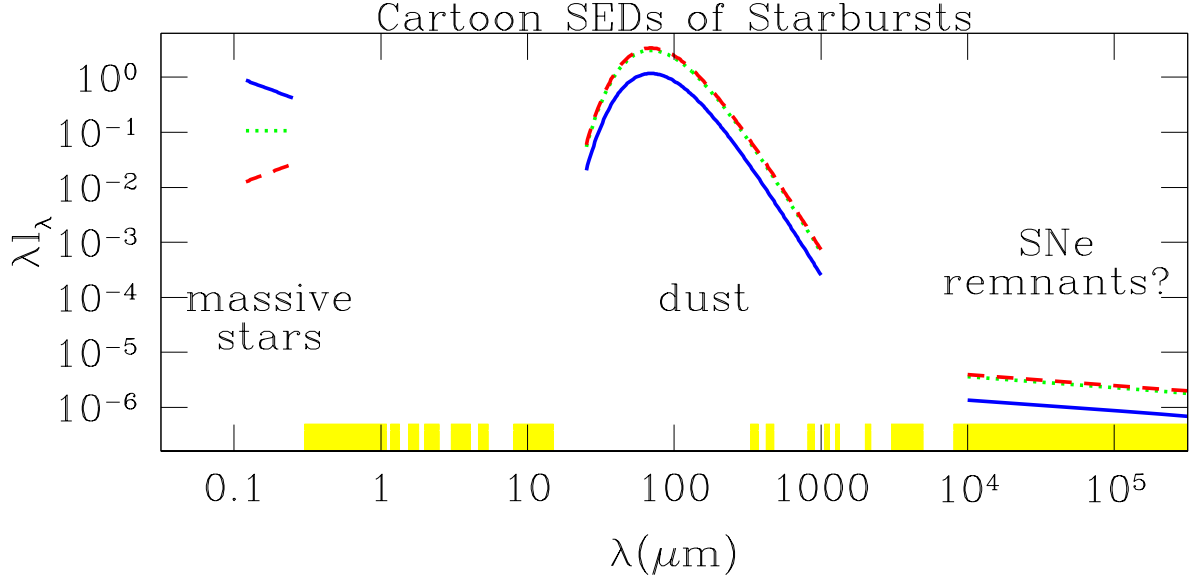}
\caption{Schematic spectra of a starburst with fixed star-formation rate
and varying dust opacity.  These spectra illustrate
the empirical correlations between UV, far-IR, and radio properties
of actively star-forming galaxies in the local universe.
Increasing the dust obscuration makes
the UV continuum redder and fainter while boosting the 
far-IR luminosity; increases in far-IR luminosity are
accompanied by increases in radio luminosity.
The range of obscurations shown ($A_{1600}\sim 0.5$, solid line;
$A_{1600}\sim 2.5$, dotted line; $A_{1600}\sim 4.5$, dashed line) roughly
span the range observed in local starbursts.  At $A_{1600}\sim 2.5$
most of the luminosity of massive stars is absorbed by dust.
Further increases in dust opacity do not significantly increase
the absorbed energy, and so the dotted and dashed lines
nearly overlap in the far-IR and radio.
Dust emission in the far-IR is typically the dominant component
in rapidly star-forming galaxies bolometric luminosities,
but most of this emission cannot be detected through
atmospheric windows (shaded regions on the lower abscissa).
}
\end{figure}

\newpage
\begin{figure}
\figurenum{2}
\plotone{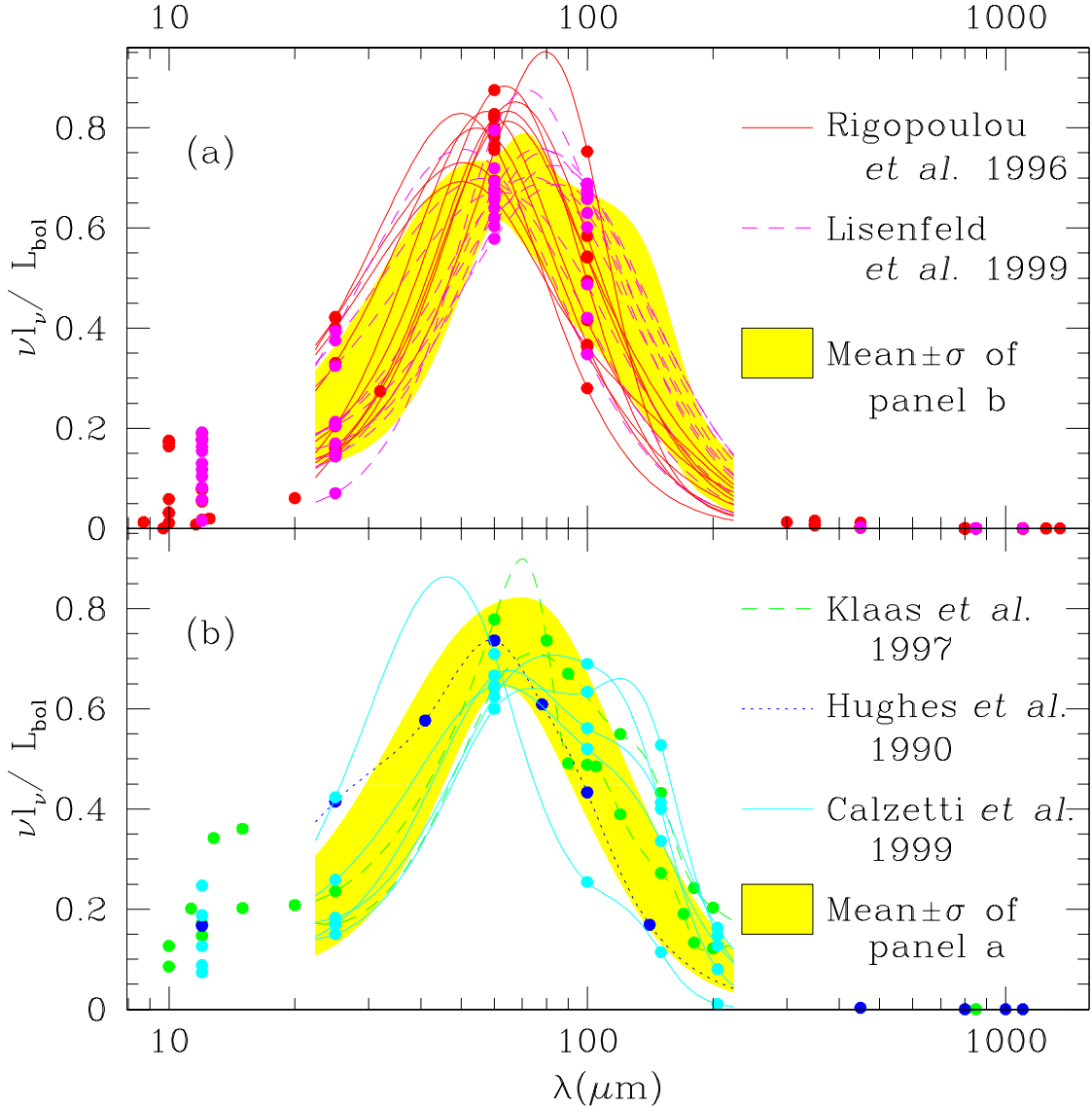}
\caption{
The far-IR spectral energy distributions of plausible local analogs
to rapidly star-forming high-redshift galaxies.
The points show measurements (see text for references) and the solid
lines spline fits to these measurements.  Each SED has been scaled to
have the same total luminosity within the range of the spline fit.
Panel (a) shows the dust SEDs of galaxies with unusually
large IR luminosities (LIRGs and ULIRGs).  Panel (b) shows
galaxies where more detailed dust SED measurements are
available; these are mainly less luminous UV-selected galaxies.
The dust SEDs of these two samples are not drastically
different, as the shaded regions show.
Because of the range in $\nu l_\nu / L_{\rm bol}$ at each wavelength,
measuring a galaxy's flux at only one far-IR/sub-mm wavelength
cannot pin down precisely its bolometric dust luminosity.
This statement is quantified in \S 2.1 and Table 1.
}
\end{figure}

\newpage
\begin{figure}
\figurenum{3}
\plotone{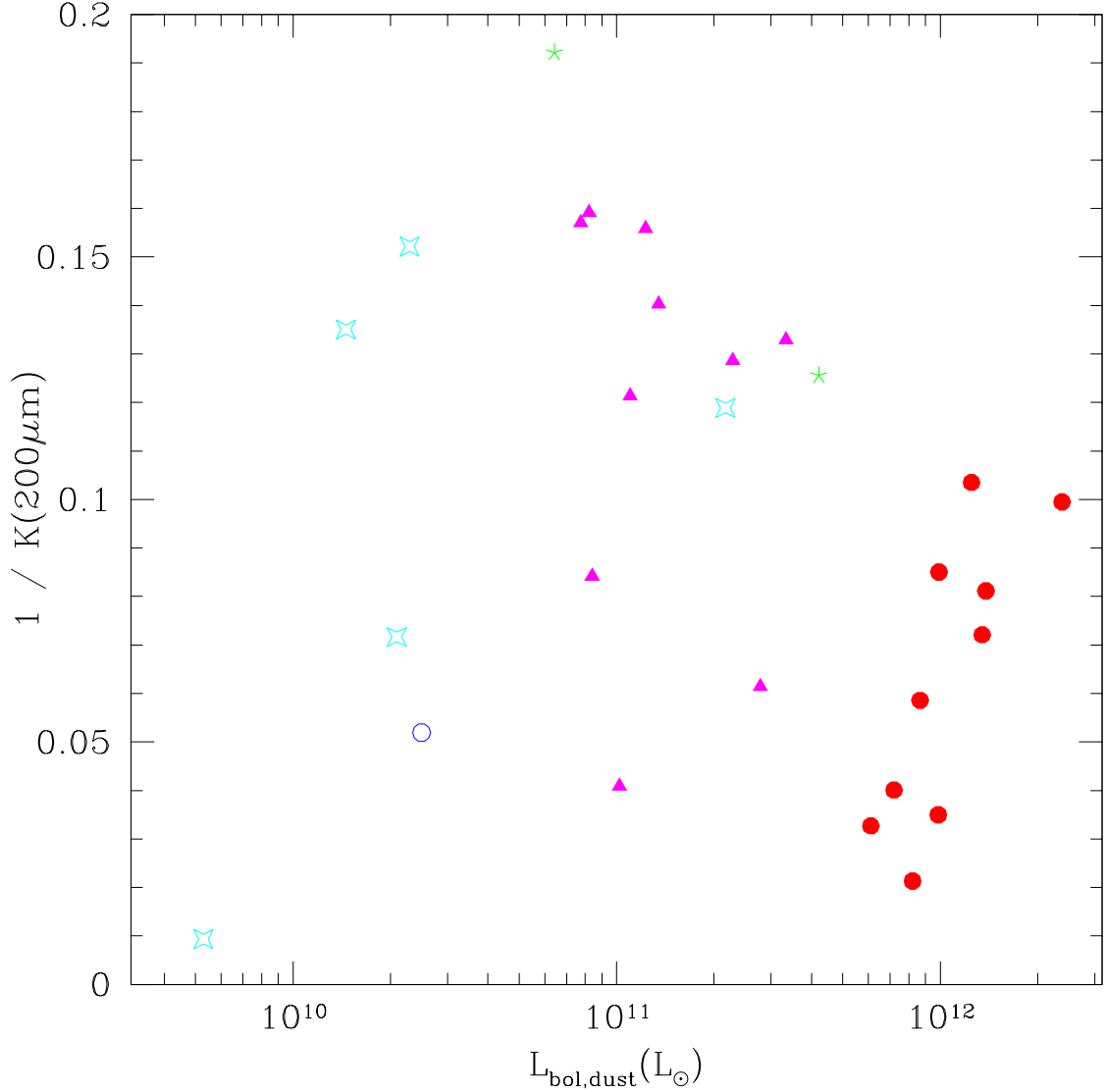}
\caption{
Dust SED shapes at different bolometric dust luminosities.
The quantity ${\cal K}(200\mu{\rm m})\equiv L_{\rm bol,dust}/\nu l_\nu(200\mu{\rm m})$ 
is sensitive to differences in the shapes of galaxies' dust SEDs;
its value will be important to us below when we discuss
$850\mu$m fluxes from galaxies at $z\sim 3$.  Shown are
the values of $1/{\cal K}(200\mu{\rm m})$ for local galaxies spanning
a range of bolometric dust luminosities:  ULIRGs (filled circles),
LIRGs (triangles), ``interacting galaxies'' (filled stars), UV-selected
starbursts (open stars), and M82 (open circle).  
More complete descriptions of these local samples are presented
in the text.  The bolometric luminosities assume that
each galaxy is at a distance of $cz / (75 {\rm km/s/Mpc})$,
except for M82 where we used 3.2 Mpc (Freedman \& Madore 1988).
Brighter galaxies tend to have hotter dust
temperatures and larger values of ${\cal K}(200\mu{\rm m})$,
but this trend has a great deal of scatter and cannot
easily be used to improve estimates of galaxies' bolometric
dust luminosities derived from observations at $\sim 200\mu$m rest.
}
\end{figure}

\newpage
\begin{figure}
\figurenum{4}
\plotone{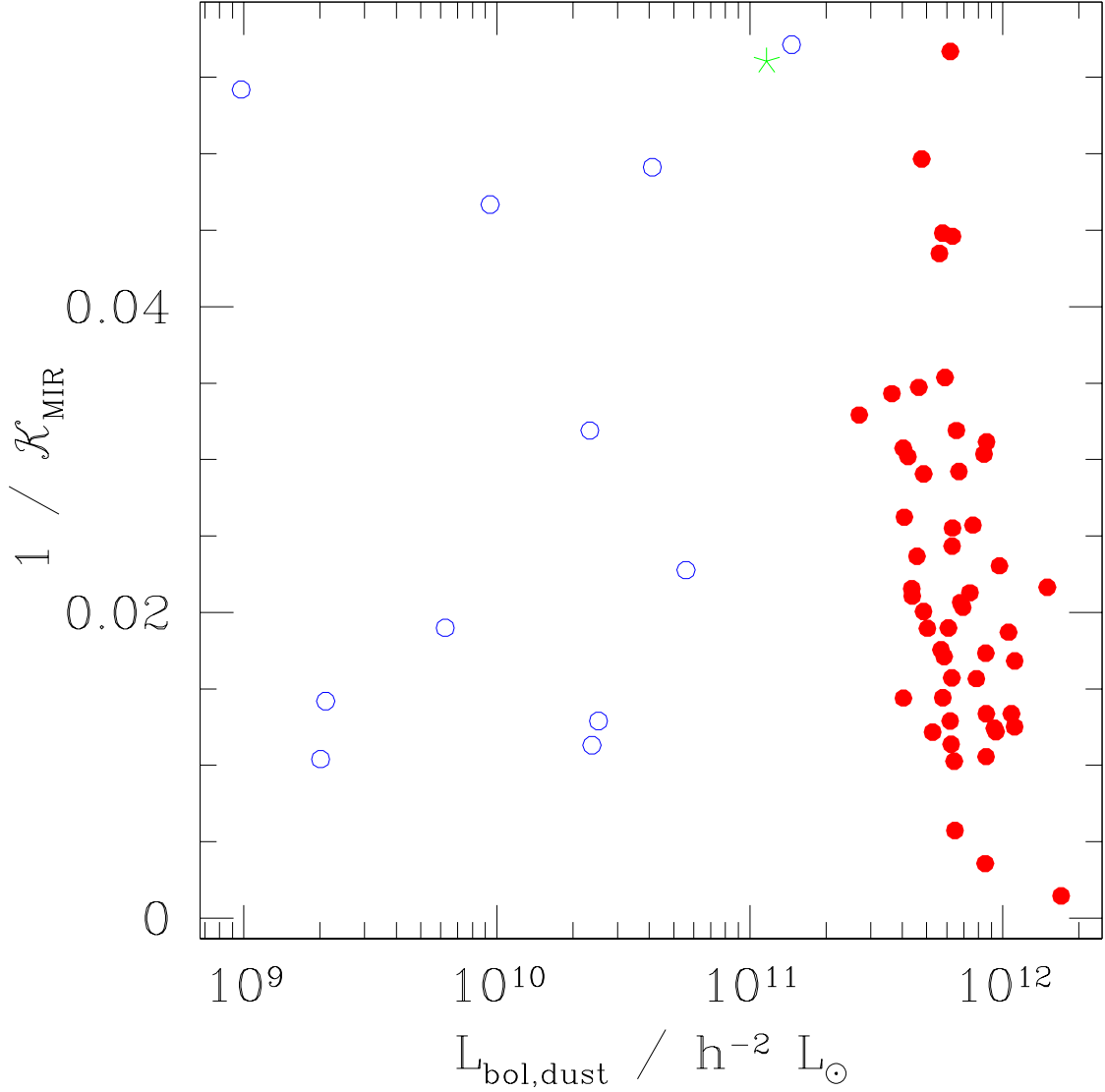}
\caption{
The ratio of $6<\lambda<9\mu$m PAH luminosity to bolometric
dust luminosity 
(${\cal K}_{\rm MIR}\equiv L_{\rm bol,dust}/\int_{6\mu{\rm m}}^{9\mu{\rm m}}d\lambda\, l_\lambda$) 
for rapidly star-forming galaxies in the local universe. 
Open and solid circles represent starbursts and star-formation--dominated ULIRGs
observed by Rigopoulou \et (2000).  The star represents the starburst NGC6090
observed by Acosta-Pulido \et (1996).  Similar values
of $K_{MIR}$ are observed in rapidly star-forming galaxies
with a large range of bolometric luminosities.
}
\end{figure}

\newpage
\begin{figure}
\figurenum{5}
\plotone{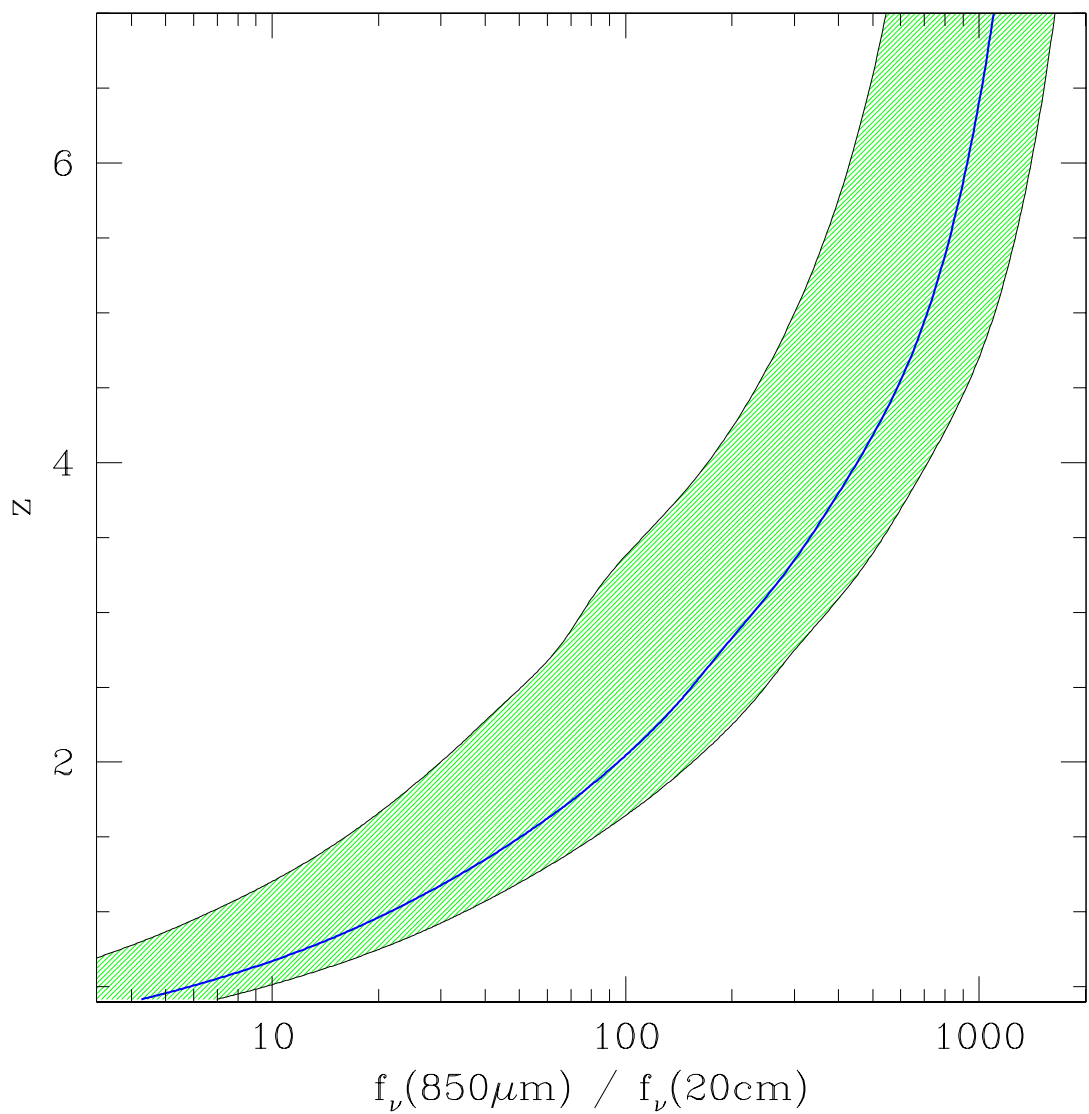}
\caption{
The ratio of sub-mm to radio flux as a redshift indicator.  
The shaded region shows $1\sigma$ limits on the ratio
$f_\nu(850\mu{\rm m}) / f_\nu(20{\rm cm})$ for rapidly
star-forming galaxies at different redshifts; the solid
line shows the mean ratio.  The error bars are
skewed for reasons discussed in \S 2.4.
See also Carilli \& Yun (1999, 2000).
}
\end{figure}

\newpage
\begin{figure}
\figurenum{6}
\plotone{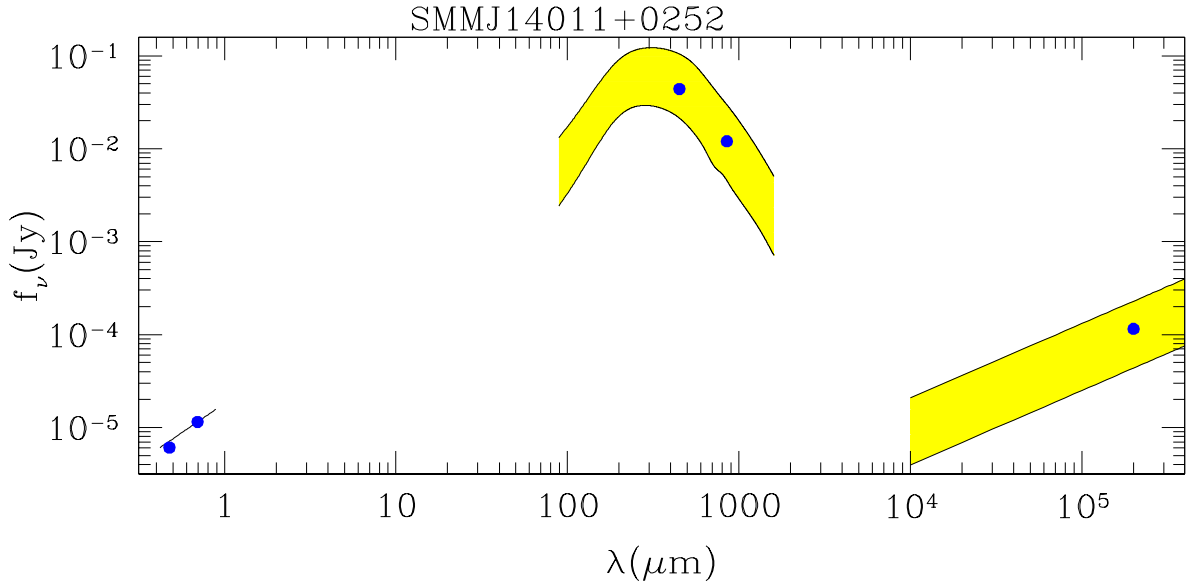}
\caption{
The optical to radio SED of the lensed star-forming galaxy
SMMJ14011+0252 at $z=2.565$.  
Circles show measured fluxes (see text for references).
$G$ and ${\cal R}$ photometry were used to estimate
the slope of this galaxy's (rest) far-UV continuum (solid line).
This slope was then used to predict the far-IR and
radio properties of SMMJ14011 as described in
sections 2 and 3.1.  The measured sub-mm and radio fluxes
fall within the $\pm 1\sigma$ predicted region (shaded),
suggesting that SMMJ14011 follows the same relation
between its UV, far-IR, and radio properties
as local starburst galaxies.
}
\end{figure}

\newpage
\begin{figure}
\figurenum{7}
\plotone{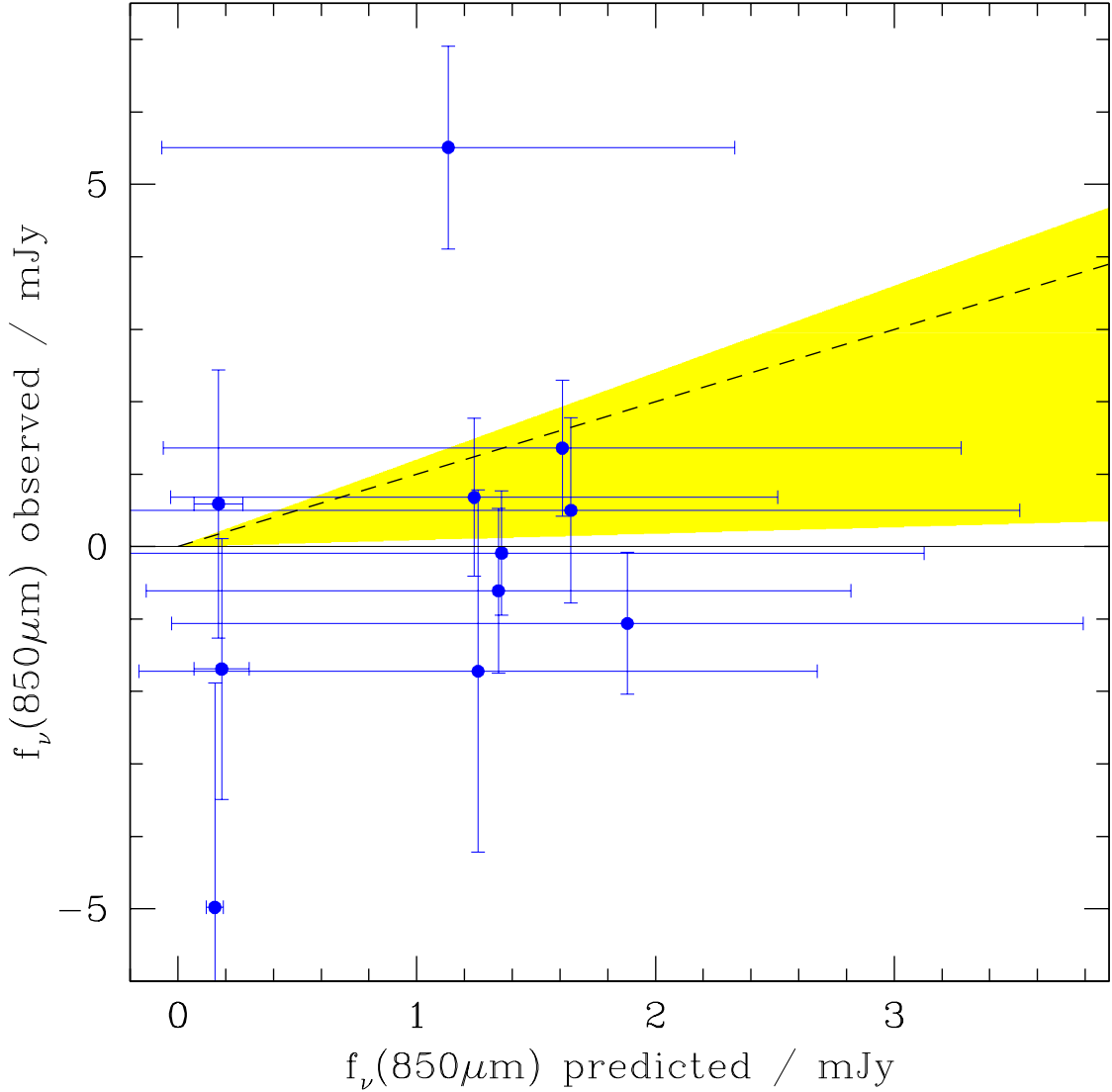}
\caption{
The predicted and observed $850\mu$m fluxes of 11 LBGs observed
by Chapman \et (2000).  The shaded region shows the 90\% credible
interval on the slope of the line 
$f_{850,{\rm observed}} = b\times f_{850,{\rm predicted}}$.
The large error bars on the
predicted fluxes, which arise for reasons discussed in \S 3.2,
are plotted symmetrically about the mean for simplicity, though
in fact they are very skewed.  Monte-Carlo simulations suggest
that most Lyman-break galaxies will have observed fluxes
smaller than the predicted mean and that a small fraction will have
observed fluxes significantly larger.
}
\end{figure}

\newpage
\begin{figure}
\figurenum{8}
\plotone{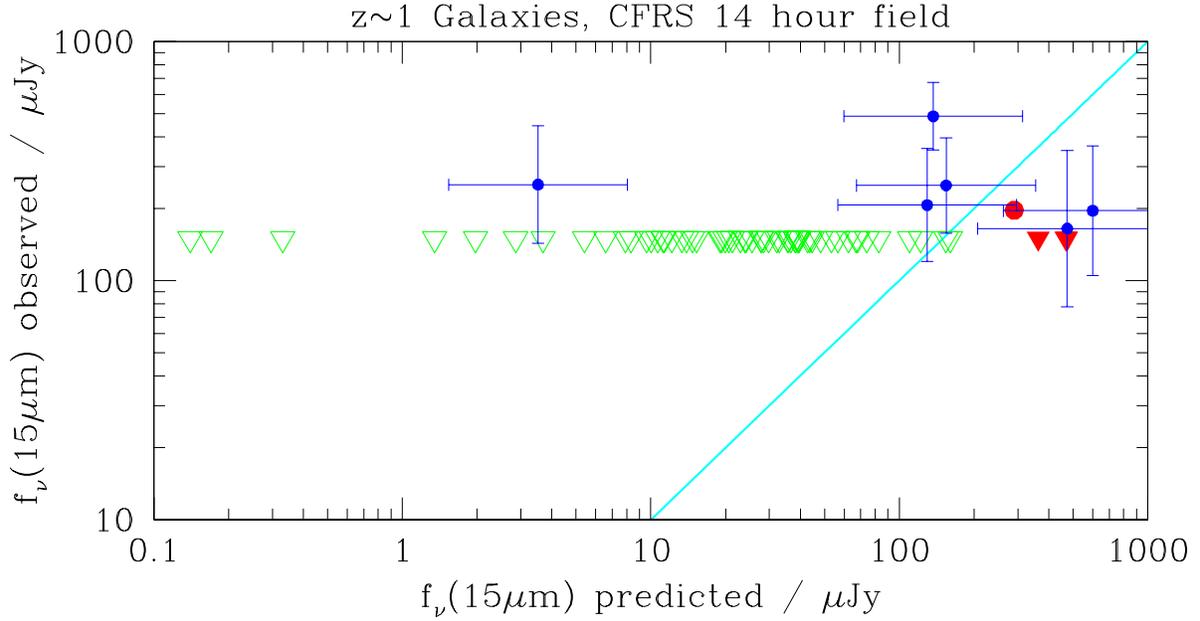}
\caption{
The predicted and observed $15\mu$m fluxes of Balmer-break
galaxies at $z\sim 1$.  The majority of these galaxies
are undetected at a $3\sigma$ limit of $150\mu$Jy;
uncertainties on the predicted fluxes of these galaxies have
been omitted for clarity, and the upper limits to their
observed fluxes are indicated with downward pointing triangles.
The six detections are indicated by points with
error bars in both $x$ and $y$; the one
with $f_{15,{\rm predicted}}\ll f_{15,{\rm observed}}$
has a significant separation between its optical
and mid-IR centroids and may be a misidentification.
Three Balmer-break galaxies in this field are too red in
the far-UV to satisfy the active star-formation criterion
discussed in \S 3.3; their locations on this plot
are indicated by smaller triangles for the non-detections
and by a point without error bars for the detection.
}
\end{figure}

\newpage
\begin{figure}
\figurenum{9}
\plotone{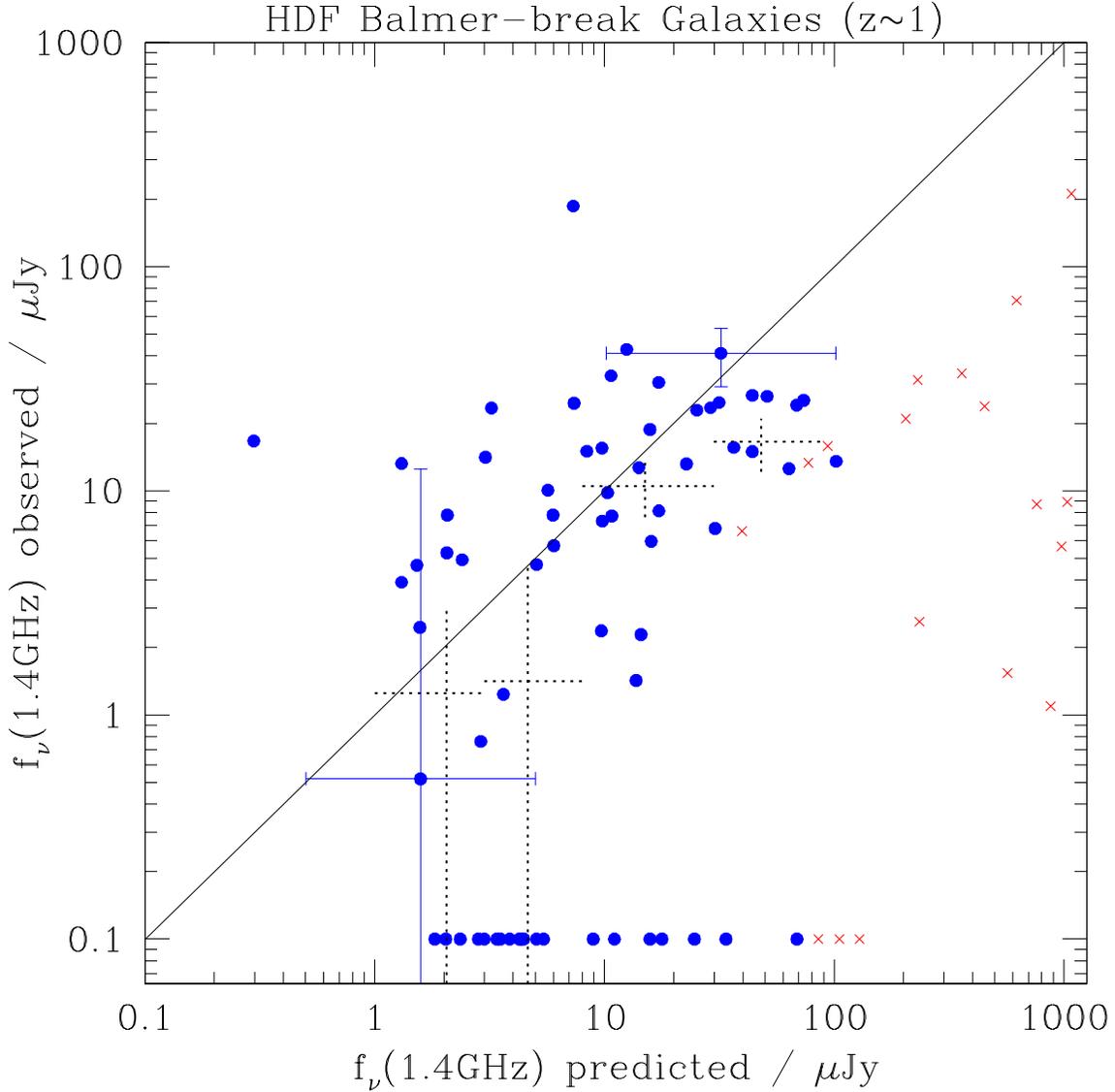}
\caption{
The predicted and observed 20cm fluxes of Balmer-break
galaxies at $z\sim 1$.  For clarity uncertainties are
shown for only a few representative points.  Galaxies
with observed fluxes lower than $0.1\mu$Jy are shown
on the plot at $0.1\mu$Jy.  The locations
of Balmer-break galaxies too red to satisfy the
active star-formation criterion of \S 3.3 are indicated with
solid crosses.  The larger dotted crosses indicate
the mean observed flux and standard deviation of the mean
for galaxies in different bins of predicted flux.
One outlier was excluded from the calculation
of the mean and standard deviation in its bin
of predicted flux, the galaxy with $f_{\rm pred}\simeq 7.5$, 
$f_{\rm obs}\simeq 200$, which likely has a significant AGN
contribution to its measured flux.
}
\end{figure}

\newpage
\begin{figure}
\figurenum{10}
\plotone{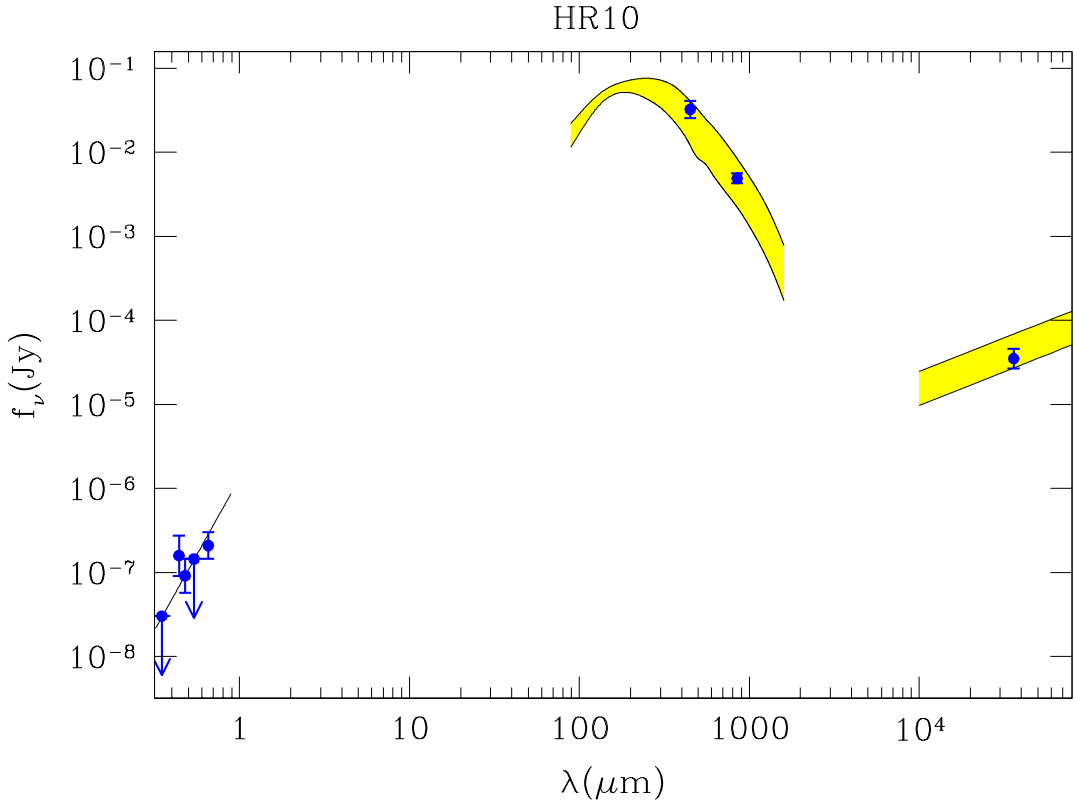}
\caption{
The far-UV to radio SED of HR10.  The shaded regions,
adjusted vertically to match the observed $450\mu$m,
$850\mu$m, and 3.6cm photometry,
show $\pm 1\sigma$ uncertainties on the
shape of this galaxy's far-IR and radio emission
under the assumption that it obeys the local
FIR/radio correlation and has a dust SED similar
to those of actively star-forming galaxies in the
local universe.  The very uncertain (rest) far-UV
photometry for this galaxy does not allow us
to estimate its far-UV spectral slope $\beta$;  the
solid line in the UV indicates the slope that
MHC's $\beta$/far-IR relation would predict.
HR10 has a far higher ratio of far-IR to far-UV
luminosity than the other galaxies we have
considered, and would not be included in
most UV-selected surveys despite its large
star-formation rate.
}
\end{figure}

\newpage
\begin{figure}
\plotone{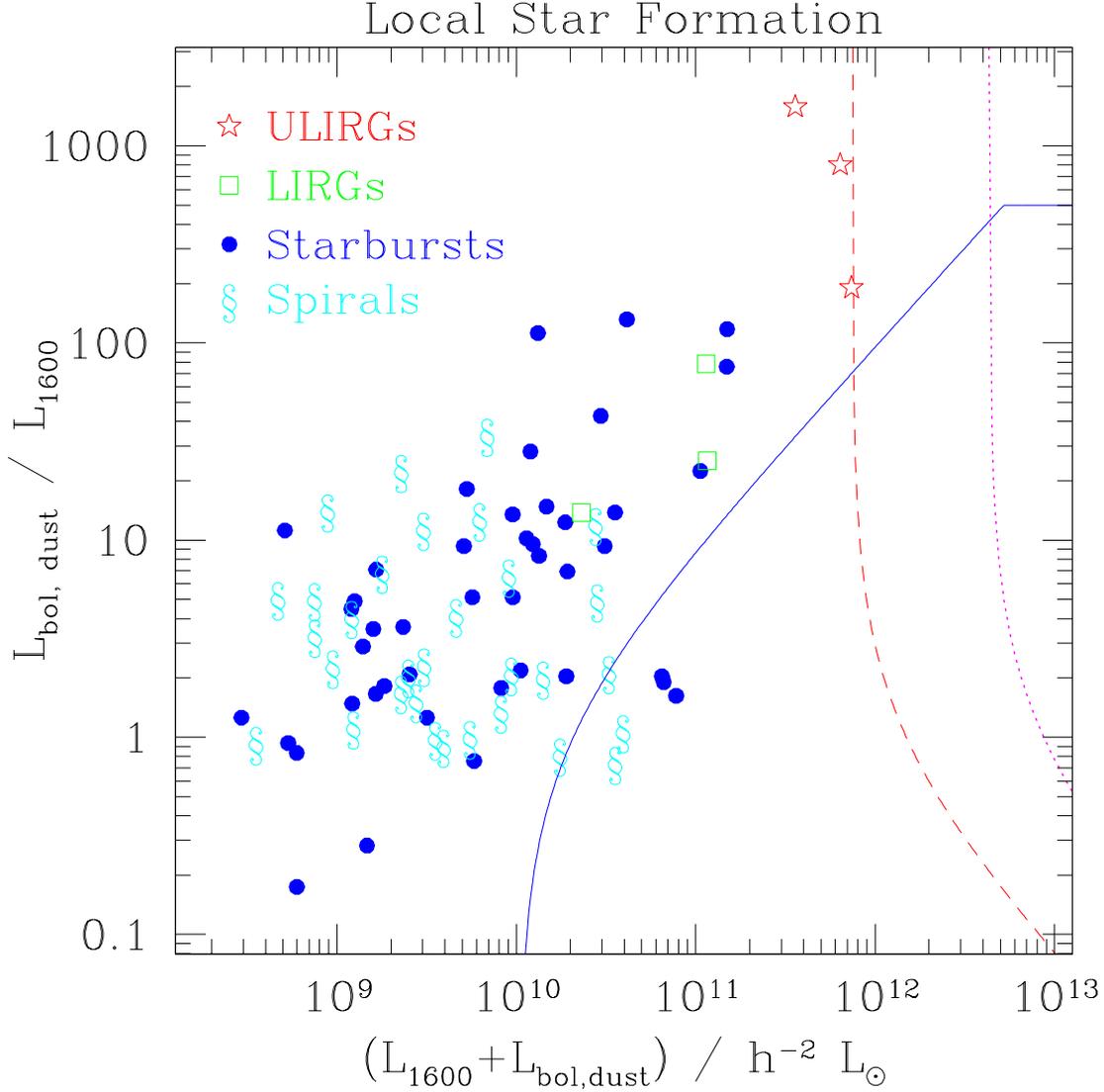}
\figurenum{11a}
\caption{Star formation in the local universe.  The sum
$L_{1600}+L_{\rm bol,dust}$, on the abscissa, provides
a crude measure of star formation rate; the ratio
$L_{\rm bol,dust}/L_{1600}$, on the ordinate, provides
a rough measure of dust obscuration.  Star formation in
the local universe occurs among galaxies with a wide
range of luminosities and dust obscuration, and
more rapidly star-forming galaxies tend to be
more heavily obscured.  See text
for a description of the four local galaxy samples
(spiral, starburst, LIRG, ULIRG) shown.  Galaxies similar to those
in this plot would not be included in existing
optical, sub-mm, or radio surveys at $z\sim 3$, which can detect
objects lying to the right of the solid, dashed, and dotted lines.
}
\end{figure}

\newpage
\begin{figure}
\plotone{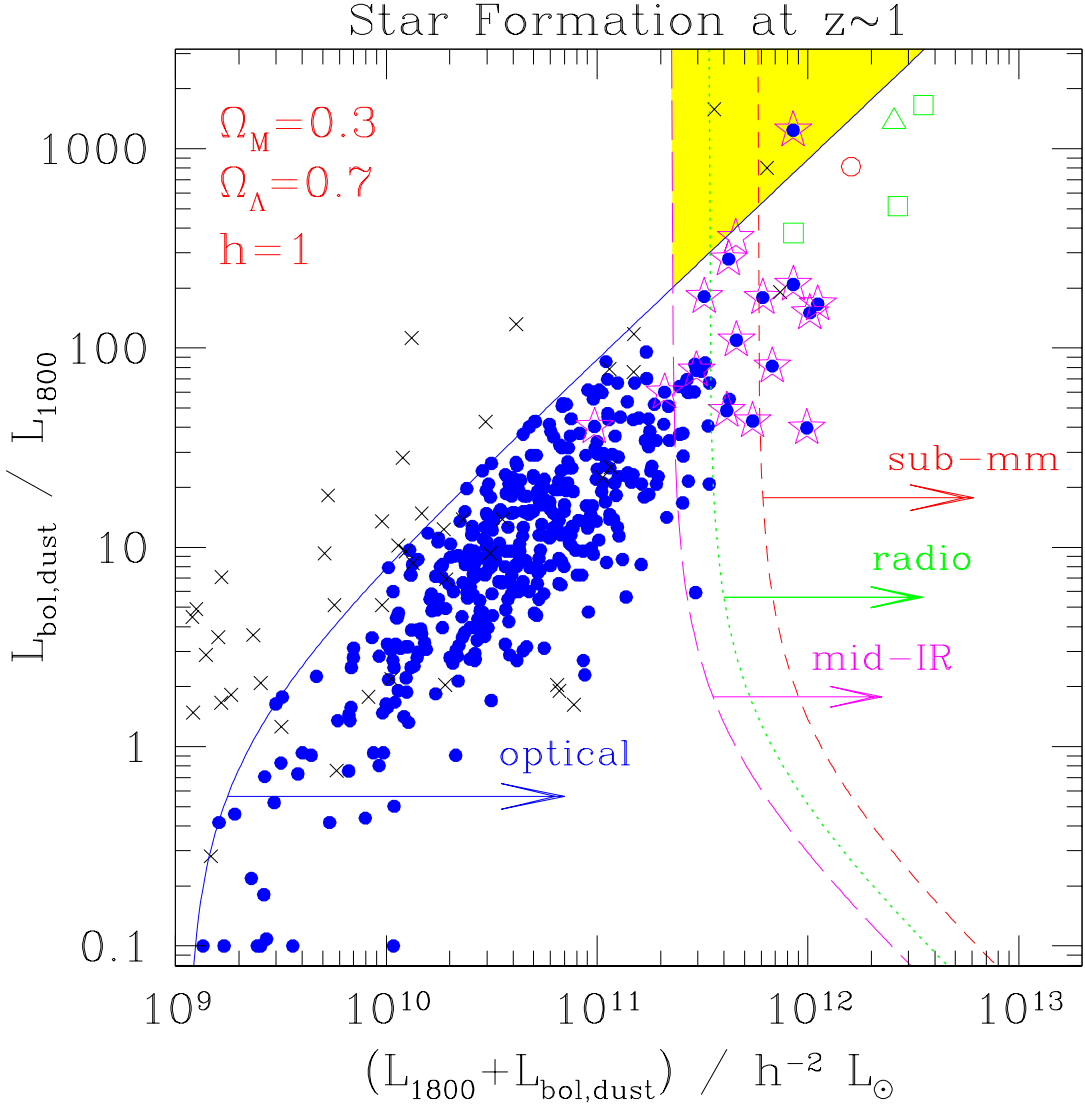}
\figurenum{11b}
\caption{Star-formation at $z\sim 1$.
The abscissa and ordinate are as in Figure 11a.  Solid circles
represent optically selected Balmer-break galaxies at $z\sim 1$,
stars represent $z\sim 1$ ISO $15\mu$m sources (15/16 of which
are also Balmer-break galaxies),
open squares (triangles) represent
$850\mu$m/radio sources with estimated redshifts $z<2$
from BCR that have detected (undetected)
optical counterparts, and the open circle
represents HR10.
Crosses mark the positions of the local starbursts, LIRGs, and ULIRGs
from Figure 11a.  Galaxies at $z\sim 1$ exhibit a similar
correlation of luminosity and dust obscuration
as local galaxies.  The lines show $z=1.0$ completeness
limits for existing surveys at various wavelengths.
Objects in the shaded region of the plot can be detected
only through their dust emission.
}
\end{figure}

\newpage
\begin{figure}
\plotone{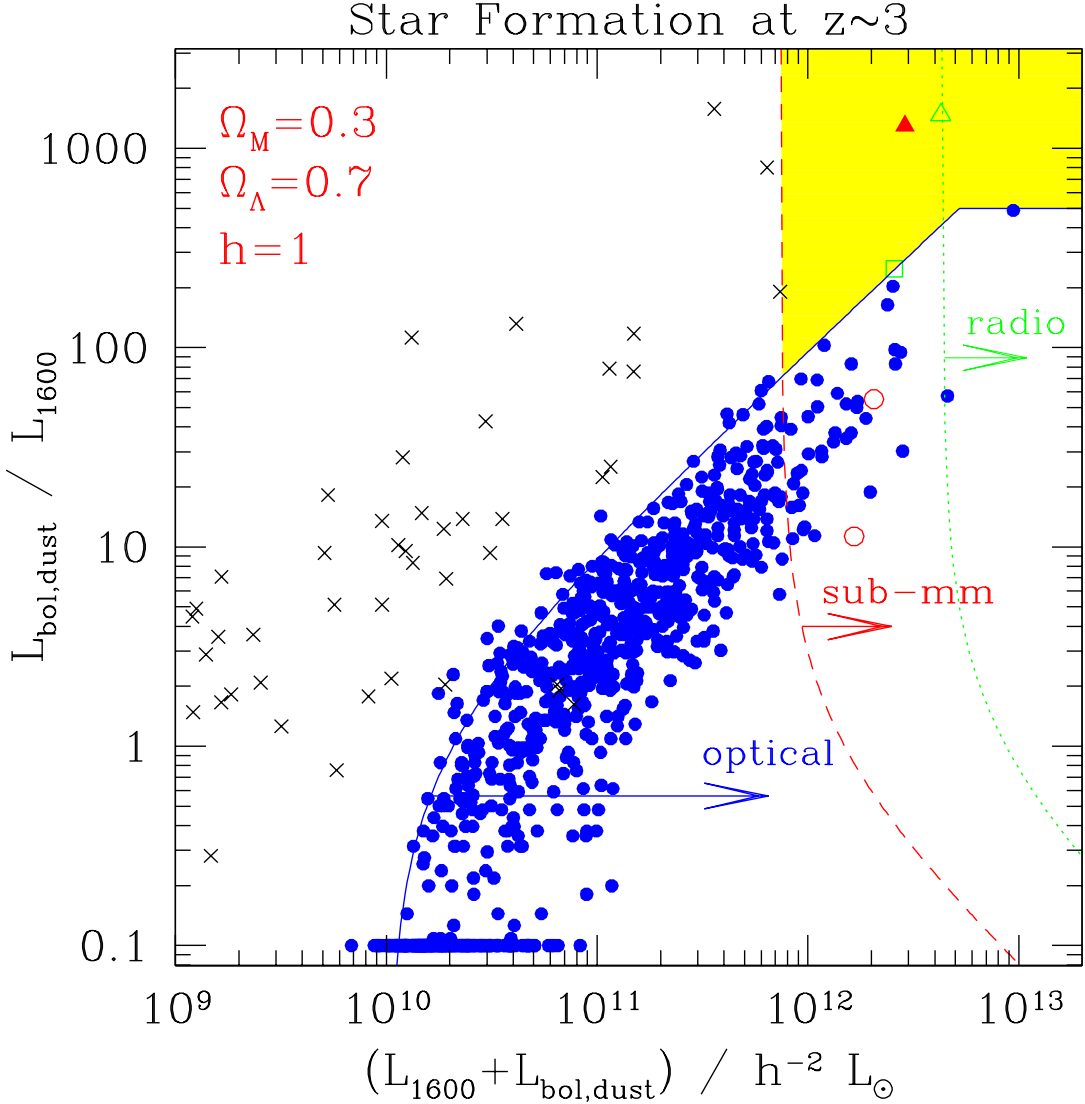}
\figurenum{11c}
\caption{Star-formation at $z\sim 3$.  The abscissa and ordinate
are as in Figure 11a.  Solid circles represent optically selected
Lyman-break galaxies, open circles represent the optically identified
$z\sim 3$ $850\mu$m sources SMMJ14011+0252 and west MMD11,
open squares (triangles) represent
$850\mu$m/radio sources with estimated redshifts $z>2$
from BCR that have detected (undetected)
optical counterparts, and the solid triangle represents
the optically undetected sub-mm source SMMJ00266+1708.
Crosses represent local starbursts, LIRGs, and ULIRGs.
Solid, dashed, and dotted
lines show the approximate $z=3.0$ completeness limits of existing
optical, sub-mm, and radio surveys.
Besides a 1600\AA\ flux limit, the optical line assumes
a $G-{\cal R}$ color limit of 1.2; the probability that
a $z\sim 3$ galaxy will be blue enough to satisfy standard LBG selection
criteria declines steadily for $L_{\rm bol,dust}/L_{1600}\simgt 20$
and reaches zero by $L_{\rm bol,dust}/L_{1600}\sim 500$---assuming
that these galaxies obey MHC's $\beta$/far-IR relationship.
See Steidel \et (1999) and Adelberger \et (2000) for a more complete discussion.
Any galaxies in the shaded region of this diagram would be detectable
in blank-field sub-mm surveys but not in standard
optical surveys.
}
\end{figure}

\newpage
\begin{figure}
\plotone{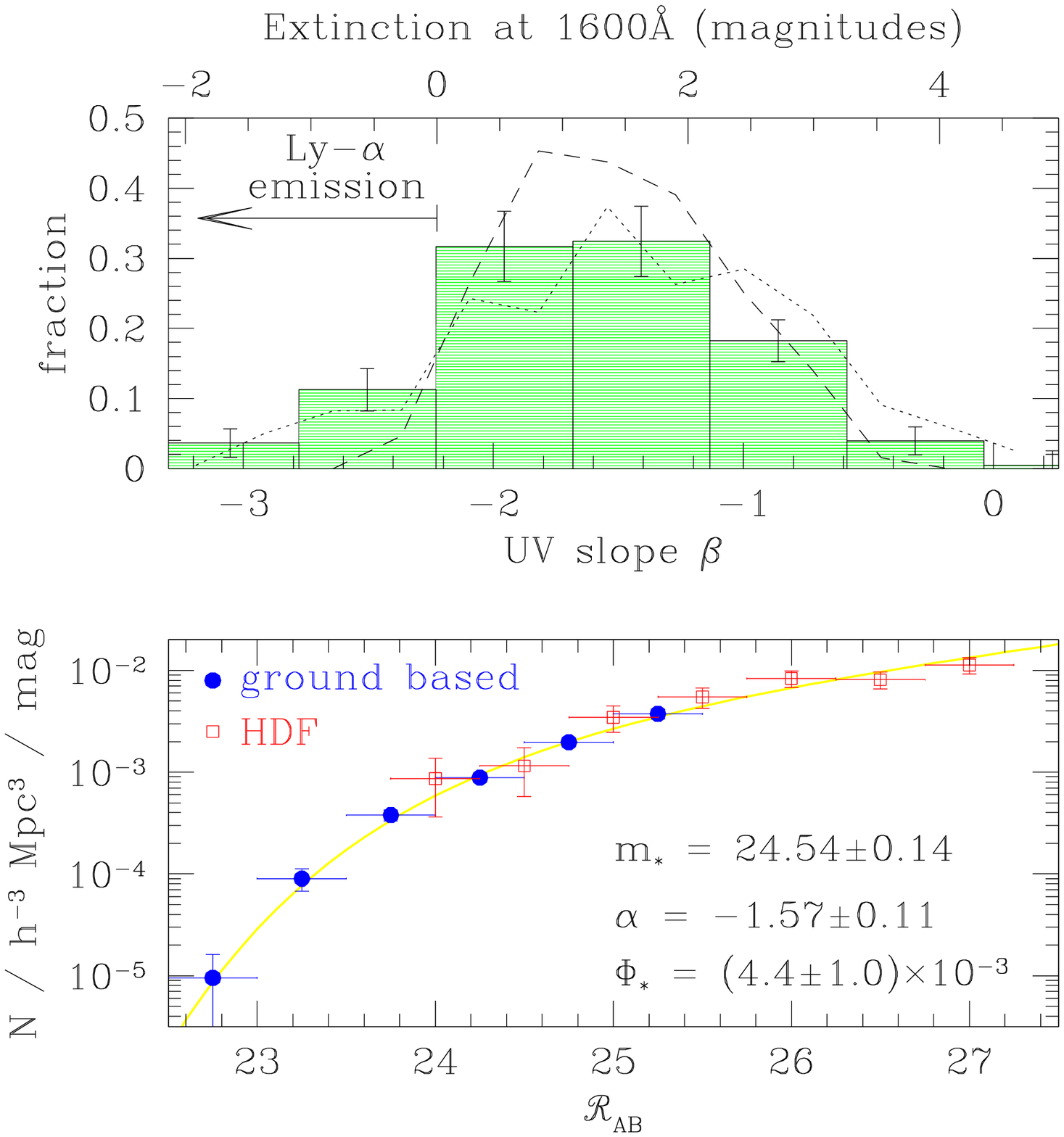}
\figurenum{12}
\caption{The apparent magnitude and $\beta$ distributions
of Lyman-break galaxies at $z\sim 3$, from Steidel \et (1999)
and Adelberger \et (2000).  The shaded $\beta$ distribution
is our current best estimate.  Systematic uncertainties
is the shape of this distribution are probably larger
than the random uncertainties denoted by the
error bars.  This is illustrated by the dotted and dashed lines, which
show, respectively, a previous estimate of the $\beta$ distribution
(derived by Steidel \et (1999) with a crude treatment of photometric errors
and incompleteness) and the $\beta$ distribution derived by
correcting for spectroscopically observed Lyman-$\alpha$ emission
but not for our photometric errors and incompleteness.
The small differences between these distributions
in their red tails have significant consequences; see below.
The number density in the
luminosity function assumes $\Omega_M=0.3$ and $\Omega_\Lambda=0.7$.
%Galaxies responsible for the $\beta<-2.2$ tail of
%the $\beta$ distribution have broad-band colors (and
%hence $\beta$ estimates) contaminated by strong Lyman-$\alpha$
%emission; these galaxies are assigned $\beta=-2.2$, corresponding
%to $L_{\rm bol,dust}=0$, in our treatment below.  
}
\end{figure}

\newpage
\begin{figure}
\figurenum{13}
\plotone{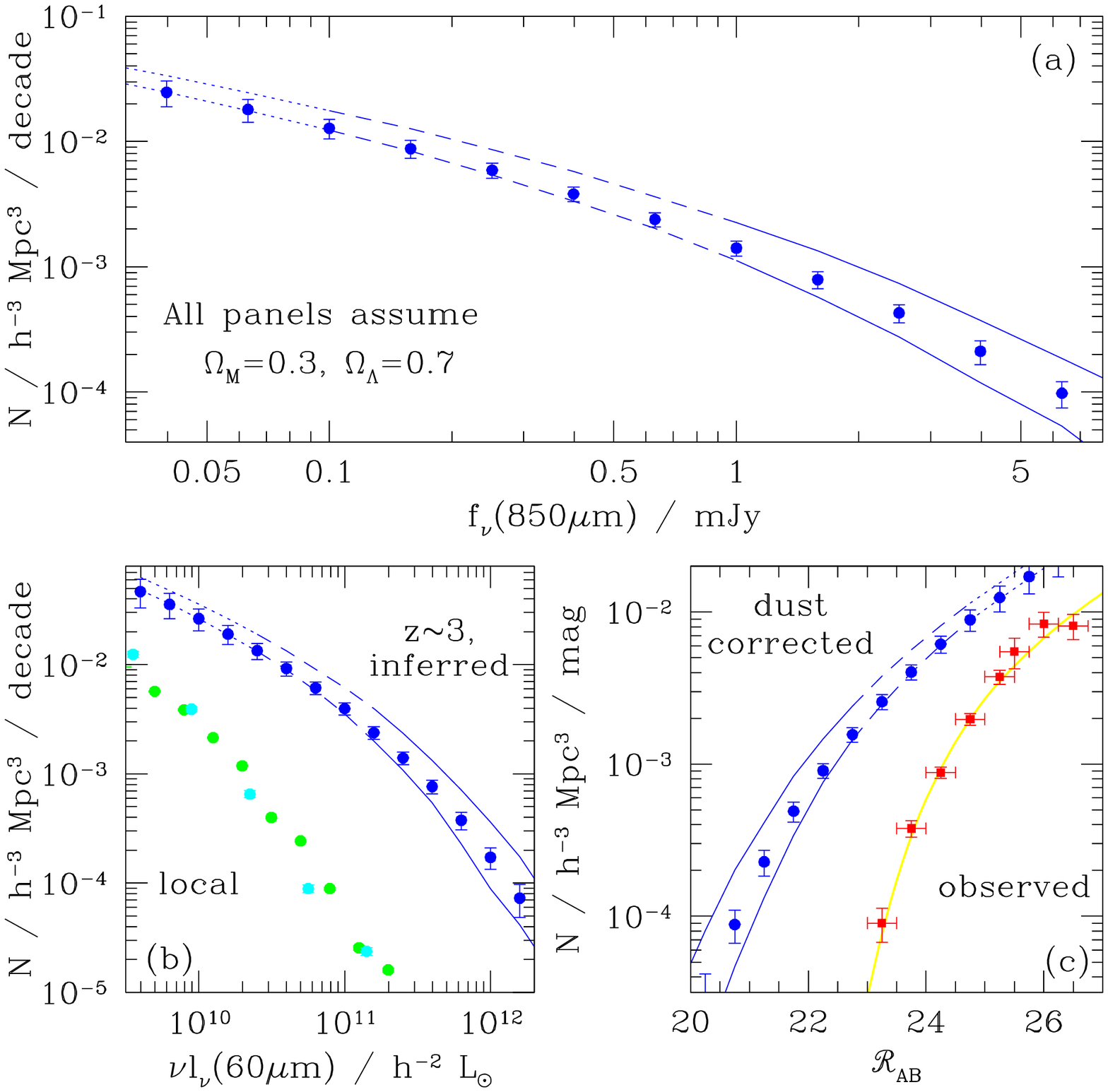}
\caption{(a)  The $850\mu$m flux distribution of
Lyman-break galaxies as inferred from their UV properties.
The points in the flux distribution
our best estimate; their error bars include the random uncertainties
discussed in the text.
The curved envelope surrounding these points shows
the much larger uncertainty in the $850\mu$m flux
distribution due to possible systematic errors in
our derived $\beta$ distribution.  At the faintest
fluxes (dotted envelope) the expected contribution
from sources with ${\cal R}>27$ accounts
for most of the estimated $850\mu$m flux distribution.
At slightly higher fluxes (dashed envelope) the
contribution from sources with ${\cal R}>25.5$ dominates.
The inferred flux distribution is most secure at
the brightest fluxes (solid envelope) that are
dominated by the contribution from galaxies
with ${\cal R}<25.5$.
(b) Comparison of the local $60\mu$m luminosity function
(Lawrence \et 1986; Soifer \et 1987) 
to the inferred $60\mu$m luminosity function of Lyman-break
galaxies at $z\sim 3$.  Symbols are as in (a). (c) The observed
and dust corrected far-UV luminosity functions of
Lyman-break galaxies at $z\sim 3$.  The observed luminosity
function is from Steidel \et (1999); the dust corrected
luminosity function was estimated as described in the text.
Symbols are as in (a).
}
\end{figure}

\newpage
\begin{figure}
\figurenum{14}
\plotone{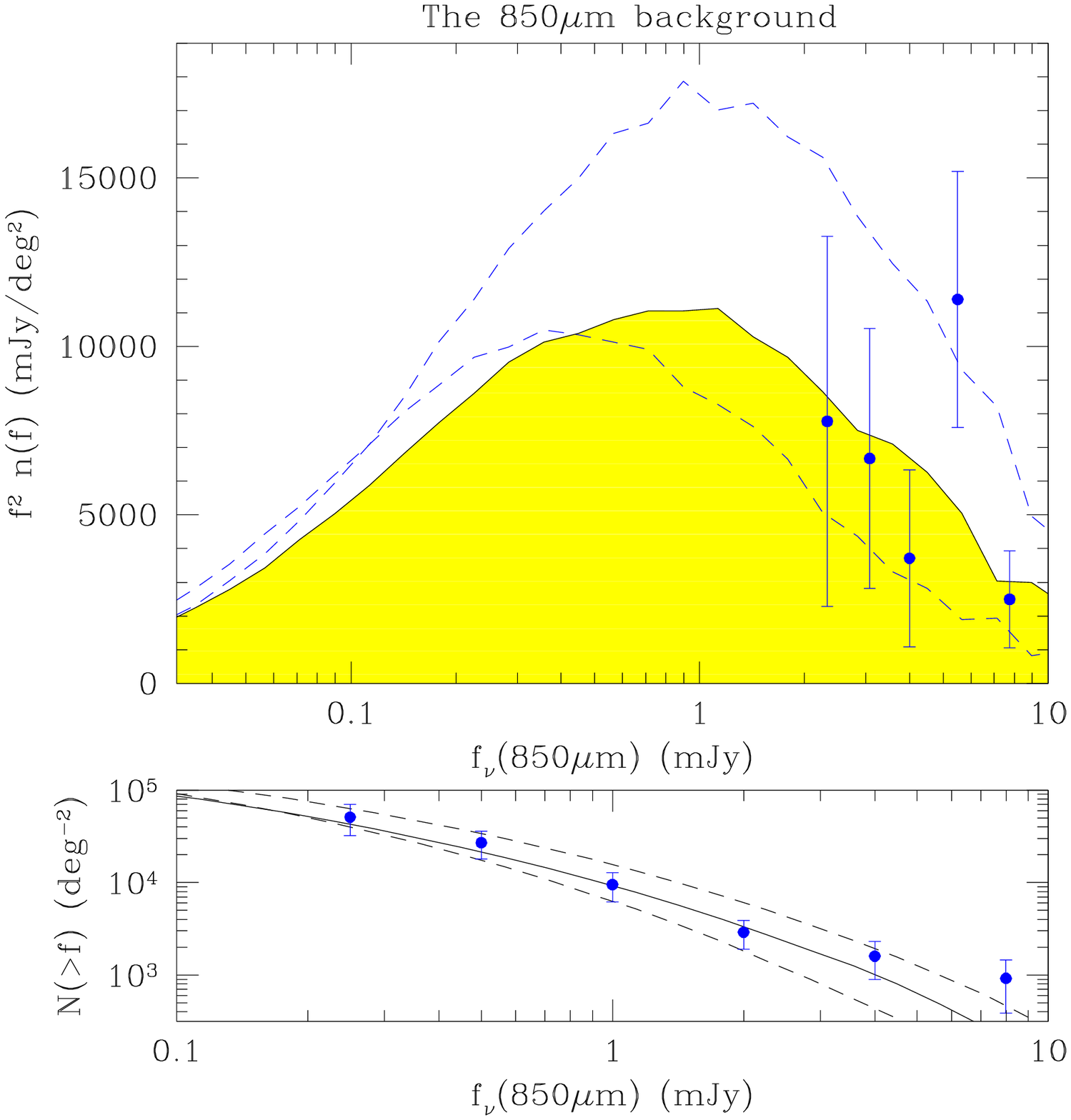}
\caption{
Top panel: The $850\mu$m background contribution $f^2_\nu n(f_\nu)$ from sources
with different fluxes.  Points with error bars show observed values,
derived from data in Barger, Cowie, and Sanders (1999); the smooth
curves show the background known UV-selected populations
at $1\simlt z\simlt 5$ would produce given the simple assumptions
described in the text.  The solid curve is appropriate
to our best estimate of the $\beta$ distribution.  Dashed lines
illustrate the uncertainty due to possible systematic
errors in our adopted $\beta$ distribution.  This is certainly
not the largest source of systematic uncertainty; is it merely
the easiest to quantify.  
The total background for our
best fit $\beta$ distribution (shaded region) is $4.1\times 10^4$ mJy/deg$^2$.
The measured $850\mu$m background is $4.4\times 10^4$ mJy/deg$^2$ (Fixsen \et 1998).
Bottom panel: The cumulative $850\mu$m number counts.  Symbols are
as in the top panel.  The observed number counts are from Blain \et (1999).
}
\end{figure}

% This forces it to spit out the oversized pages.
\clearpage

\newpage
\begin{figure}
\figurenum{15}
\plotone{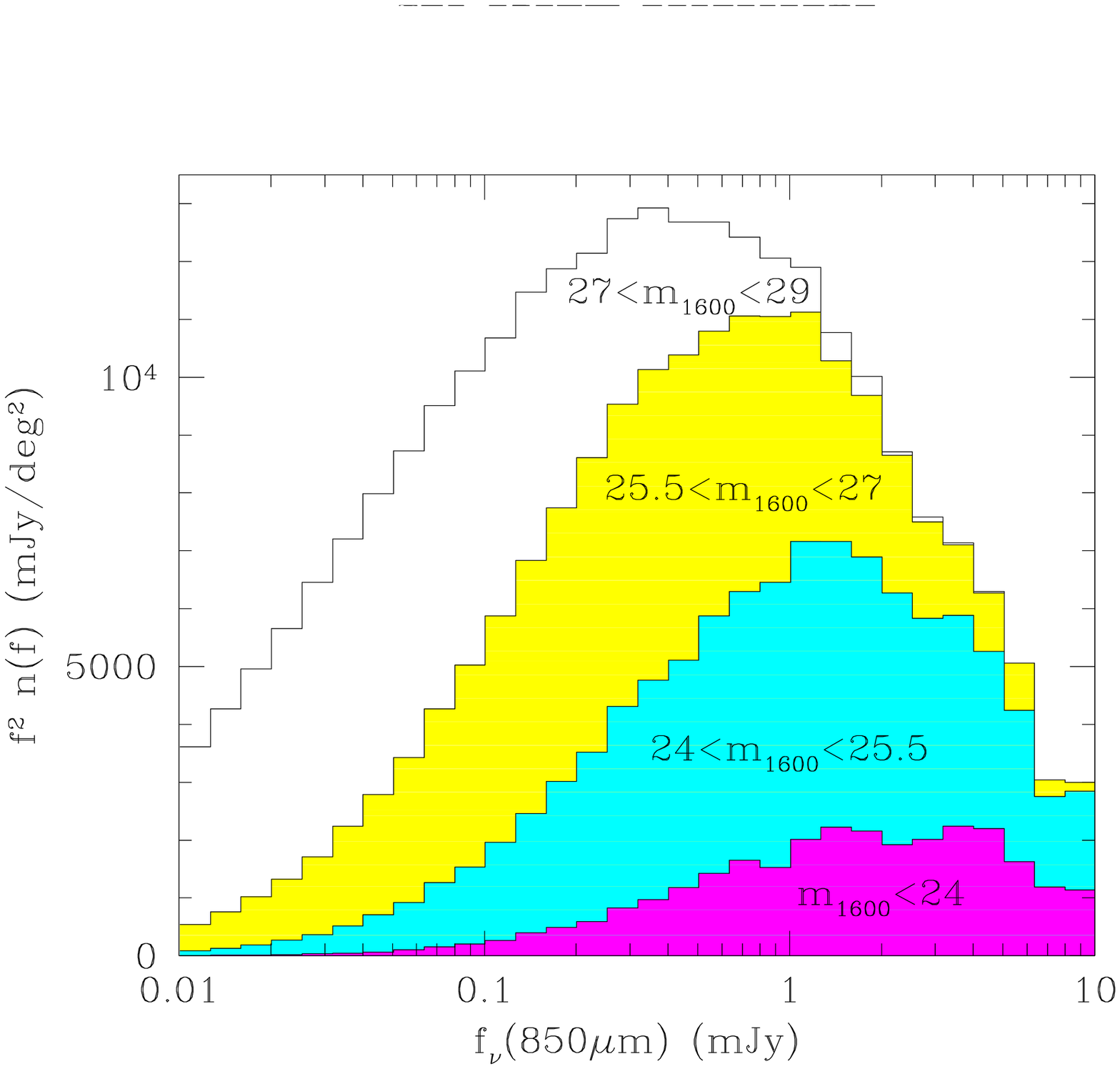}
\caption{
The expected contribution to the $850\mu$m background from UV-selected sources
with different apparent magnitudes at 1600\AA\ rest.  Even if UV-selected
populations were solely responsible for the $850\mu$m background,
we would expect most of the sub-mm sources brighter than $2$mJy at
$850\mu$m to be fainter than $m_{AB}=24$ at 1600\AA\ rest.
Also shown is the contribution to the background which would
be produced by galaxies with $27<m_{AB}<29$ if the $\alpha=-1.6$
faint end slope of the assumed luminosity function continued
past the faintest observed magnitude of $m_{1600}\simeq 27$.  These galaxies
were excluded from the calculation whose results are shown in Figure 14;
see text. 
}
\end{figure}

\begin{figure}
\figurenum{16}
\plotone{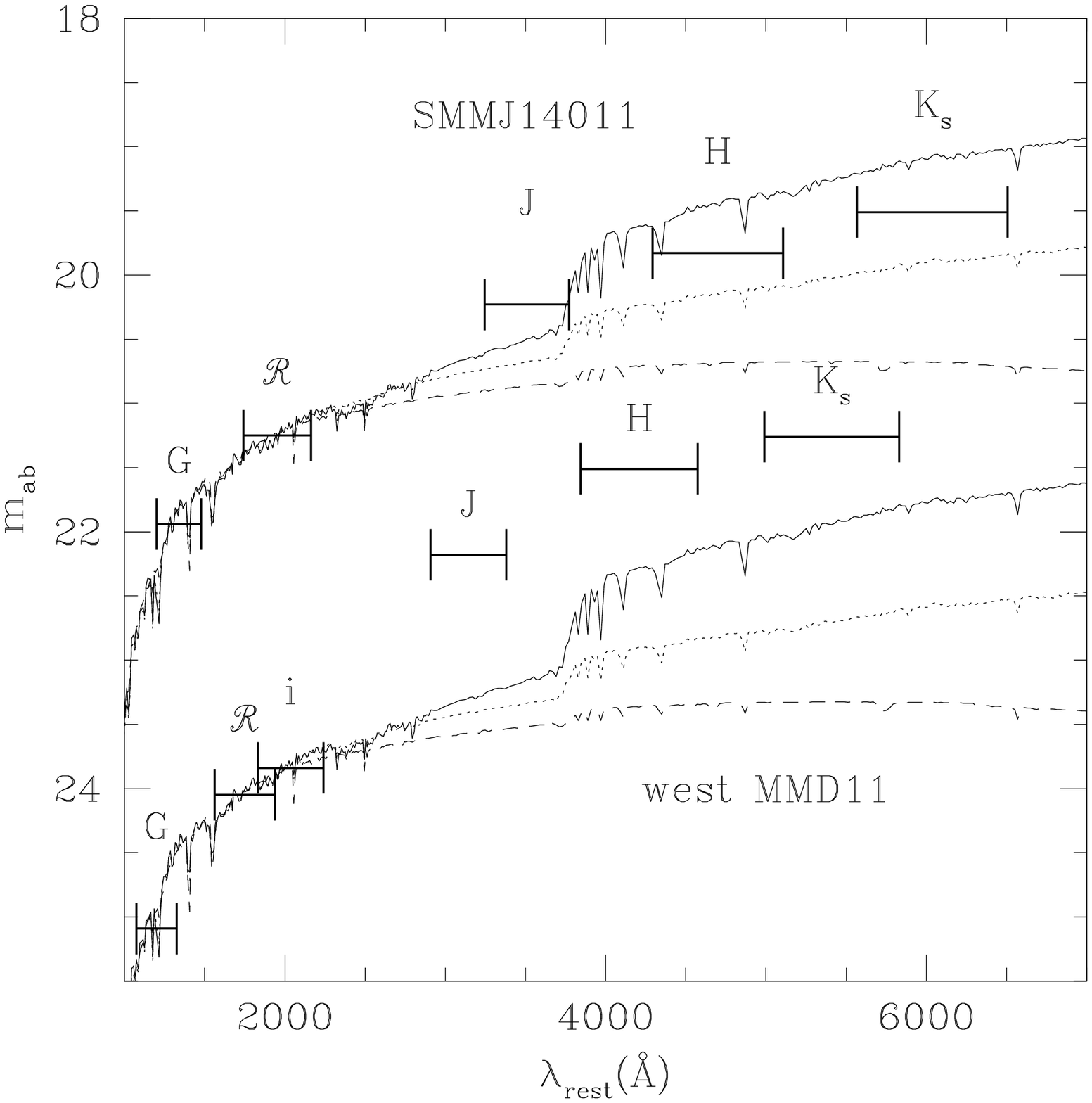}
\caption{The UV to optical photometry of two $z\sim 3$ galaxies detected
at $850\mu$m.  The bars show measured AB magnitudes, and the lines
Bruzual \& Charlot (1996) model galaxies after 1Myr (dashed), 
40 Myr (dotted), and 1Gyr (solid) of star formation at a constant rate.
The model SEDs were reddened with dust following Calzetti's (1997)
extinction law as required to match the observed $G-{\cal R}$ color;
for both galaxies this required $A_{1600}=3.2$ for the 1Gyr SED,
$A_{1600}=3.9$ for the 40Myr SED, and $A_{1600}=4.5$ for the
1Myr SED.  The value of $A_{1600}$ adopted in the text
was derived from the empirical correlation between
UV spectral slope $\beta$ and the ratio $L_{\rm bol, dust}/L_{1600}$,
not from an assumed SED and dust reddening law.
}
\end{figure}

\newpage
\begin{figure}
\plotone{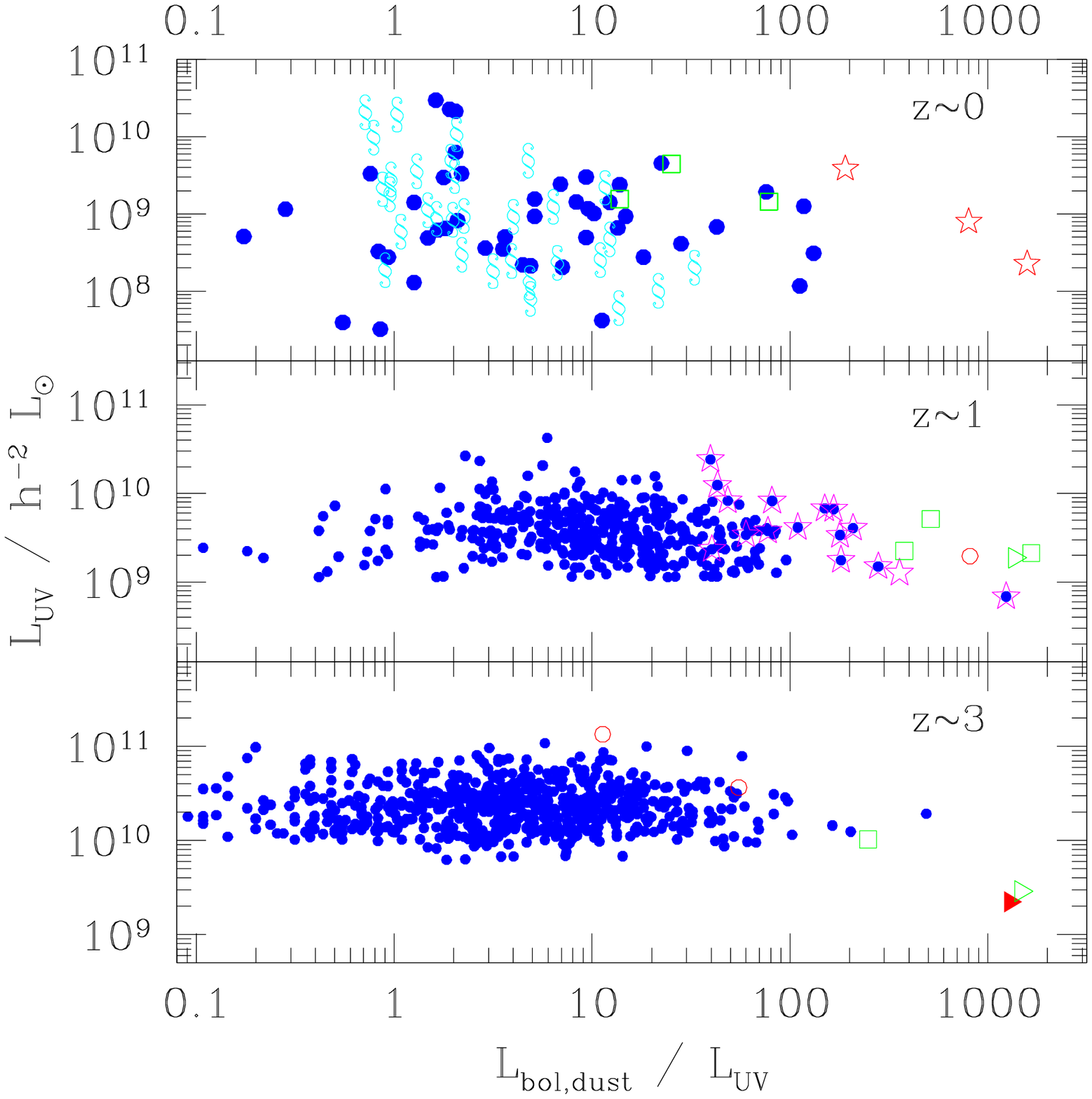}
\figurenum{17}
\caption{Far-UV luminosity versus dust obscuration for star-forming galaxies
at $z\sim 0$ (top panel), $z\sim 1$ (middle panel), and $z\sim 3$ (bottom
panel).  Symbols are as in Figures 11a--c.
Dust absorbs UV photons, but dustier galaxies
tend to have higher star-formation rates and larger intrinsic UV luminosities,
and as a result the observed UV luminosities of star-forming galaxies
are largely independent of their dust obscuration.  There are
exceptions, but in general extremely dusty galaxies
are no fainter in the far-UV than relatively dust-free galaxies.
The 1600\AA\ luminosities assume $\Omega_M=0.3$, $\Omega_\Lambda=0.7$.
Objects represented by open squares, open triangles, and solid triangles
do not have spectroscopic redshifts, and their placement on this
plot is very uncertain.
The lower limits to the UV luminosities of optically
selected galaxies at $z\sim 1$ and $z\sim 3$
reflect the flux limits of existing surveys.
The apparent increase of UV luminosities with redshift
is partly but not solely a selection effect.
The color-selection criteria used in constructing the
$z\sim 1$ and $z\sim 3$ optical samples result in a selection
bias against the bluest objects at $z\sim 1$ and the reddest
objects at $z\sim 3$.
When selection effects are corrected (as they have been
in \S 4 but not in this figure) the observed dust obscuration
distributions in optical populations at $z\sim 1$ and $z\sim 3$ are similar.
}
\end{figure}

\end{document}